\documentclass[12pt,a4paper,fleqn]{article}

\usepackage[textheight=23cm,textwidth=15cm]{geometry}
\usepackage{amsmath}
\usepackage{amssymb}
\usepackage{mathtools}
\usepackage[version=3]{mhchem}
\usepackage{graphicx}
\usepackage[american]{babel}
\usepackage{here}
\usepackage[articletitle=true, maxauthors=150]{achemso}
\usepackage{siunitx}
\usepackage[hidelinks]{hyperref}

\DeclareSIUnit{\au}{a.u.}

\begin{document}
\begin{center}

\LARGE{\bf{
       Automated Microsolvation for Minimum Energy Path Construction in Solution
}}

\large{
	Paul L. T\"urtscher\footnote{ORCID: 0000-0002-7021-5643} and
	Markus Reiher\footnote{Corresponding author; e-mail: mreiher@ethz.ch; ORCID: 0000-0002-9508-1565}
}\\[4ex]

Department of Chemistry and Applied Biosciences, ETH Zurich, \\
Vladimir-Prelog-Weg 2, 8093 Zurich, Switzerland

February 11, 2025

\vspace{0.4cm}
\bf{Abstract}
\end{center}
\vspace{-0.6cm}
{\small
Describing chemical reactions in solution on a molecular level is a challenging task due to the high mobility of weakly interacting solvent molecules which requires configurational sampling. 
For instance, polar and protic solvents can interact strongly with solutes and may interfere in reactions.
However, to define and identify representative arrangements of solvent molecules modulating a transition state is a non-trivial task. Here, we propose to monitor their active participation in the decaying normal mode at a transition state, which defines active solvent molecules. 
Moreover, it is desirable to prepare a low-dimensional microsolvation model in a well-defined, fully automated, high-throughput, and easy-to-deploy fashion, which we propose to derive in a stepwise protocol.
First, transition state structures are optimized in a sufficiently solvated quantum-classical hybrid model, which are then subjected to a re-definition of a then reduced quantum region.
From the reduced model, minimally microsolvated structures are extracted that contain only active solvent molecules. Modeling the remaining solvation effects is deferred to a continuum model. 
To establish an easy-to-use free-energy model,
we combine the standard thermochemical gas-phase model with a correction for the cavity entropy in solution.
We assess our microsolvation and free-energy models for methanediol formation from formaldehyde, for the hydration of carbon dioxide (which we consider in a solvent mixture to demonstrate the versatility of our approach), and, finally, for the chlorination of phenol with hypochlorous acid.
}

\section{Introduction}
Chemical reactions in solution can be affected by solvent molecules.
The solvent molecules can stabilize structures along a minimum energy path (MEP) through directed or isotropic (often electrostatic) interactions.
The most prominent directed interactions dominant in protic solvents are hydrogen bonds.
However, solvent molecules can also actively participate in a reaction.

The stabilizing effect of solvents on reactions
has been studied with continuum solvation models such as the polarizable continuum model (PCM)\cite{Barone1998, Tomasi2005, Mennucci2012, Herbert2021}, the conductor-like screening model (COSMO), and its refined version for 'real solvents' (COSMO-RS).\cite{Klamt2018}
As long as the solvation shell can be expected to undergo rapid configurational sampling, the averaged description of a continuum model will usually be sufficient for modeling solvation effects.\cite{Norjmaa2022}

All continuum solvation models will show limitations if strongly directional effects dominate the stabilization effect or if solvent molecules are actively involved in the reaction.
Then, so-called cluster-continuum models\cite{Pliego2020} may be employed to capture the directed stabilization effects and even active involvement of solvent molecules (for examples, see the reviews
in Refs.~\citenum{Sunoj2012, Basdogan2020a, Das2022, Norjmaa2022} and references therein).
These cluster-continuum models consider a few solvent molecules explicitly while modeling the surrounding bulk with a continuum solvation model.

However, even simulating the bulk solvation with a continuum model might be insufficient in cases where solvent molecules in an outer sphere around a reacting solute also exert directional effects.
To model such outer-sphere effects, one can employ hybrid quantum mechanics/molecular mechanics (QM/MM) approaches,\cite{Warshel1976a, Field1990, Lipparini2021, Csizi2023a} to counteract the unfavorable scaling of QM approaches with system size.
In the cluster-continuum context, the cluster is then described quantum mechanically, while the continuum is replaced with explicit solvent molecules described by a classical force field;
\cite{Demapan2022,Ho2024} see
Refs.~\citenum{Chandrasekhar2002,Yang2015,Boereboom2018,Clabaut2020} for examples and  Ref.~\citenum{Feher2019} for a comparison with continuum solvation models.

Recently, machine learning approaches have been developed to predict solvation free energies with $\Delta$-learning\cite{Meng2023} or a deep learning approach\cite{Fowles2023}, and strategies for generating reactive machine learning potentials for reactions in solution. \cite{Zhang2023}
Additionally, workflows for training neural network potentials to accelerate the sampling of the free energy surface for reactions with actively involved solvent molecules have been explored.\cite{Celerse2024}

Despite their success, cluster-continuum models as well as QM/MM models face challenges in the case of active solvent molecule participation in reactions,\cite{Sunoj2012, Norjmaa2022} because various challenges are difficult to address.
It is a priori not clear how many explicit QM solvent molecules are required to capture reaction participation effects.
Then, it is unclear where these QM solvent molecules must be placed relative to the solute.
Even if these structural challenges can be properly addressed, it is open how to define representative minimum energy paths (MEPs) of a reaction in the
presence of explicit solvent molecules since small variations in solvent molecule position produce a new MEP on the potential energy surface.
These challenges are particularly pressing in protic solvents, where proton shuffles along several molecules are possible.
Addressing these challenges often relies heavily on expert knowledge.\cite{Sunoj2012,  Varghese2019, Norjmaa2022}

With a focus on automated reaction mechanism exploration algorithms,\cite{Sameera2016, Dewyer2017, Simm2019, Unsleber2020, Steiner2022, Baiardi2022, Ismail2022, Wen2023, Margraf2023}
we need to probe
chemical reactions for active solvent involvement systematically and without the requirement of expert knowledge or manual intervention. Here, we present an automated high-throughput
pipeline that operates independently of the specific choice of solvent and explores reactions in an unbiased and automated fashion with only basic information provided as a starting point (that is, Cartesian coordinates of the reactants and a guess for the reaction coordinate that can simply consist of the reactive atoms that are supposed to approach or depart from one another, an information easily available in automated exploration procedures).
This multi-step structure preparation pipeline, called \textsc{Kingfisher}, first exploits QM/MM hybrid models to seize small solute-solvent clusters with active involvement of solvent molecules.
The clusters are then subjected to a cluster-continuum description, for which an ensemble of MEPs is generated.

The theoretical and algorithmic details of \textsc{Kingfisher} are introduced and elaborated on in Section~\ref{sec:kingf}.
In Section~\ref{sec:results},
we first study reactions in solution where polarized reactants form hydrogen bonds. Specifically, we consider the hydrolysis of formaldehyde and the chlorination of phenol in aqueous solution.
Then, we study the hydration
of \ce{CO2} in water to form carbonic acid, where \ce{CO2} acts as a hydrophobic reactant subjected to a mixture of two solvents.

\section{Microsolvated minimum energy paths}\label{sec:kingf}

The development of our pipeline follows the demands of automated chemical reaction network exploration. Aiming for minimal input information, it must
require only the Cartesian coordinates of the reactants,
information about their \textit{reactive atoms}, and the Cartesian coordinates of the solvent molecules.
Optionally, information on atoms that likely interact with solvent molecules (for instance, through hydrogen bonding) and a mixing ratio in the case of solvent mixtures could be required.
\textit{Reactive atoms} are those atoms assigned to react (for instance, according to some heuristic rule or, in a brute-force fashion, as a pair out of all possible pairs of atoms \cite{Bergeler2015,Simm2017,Grimmel2021,Unsleber2022a}).

\begin{figure}[H]
  \centering
  \includegraphics[width=0.40\textwidth]{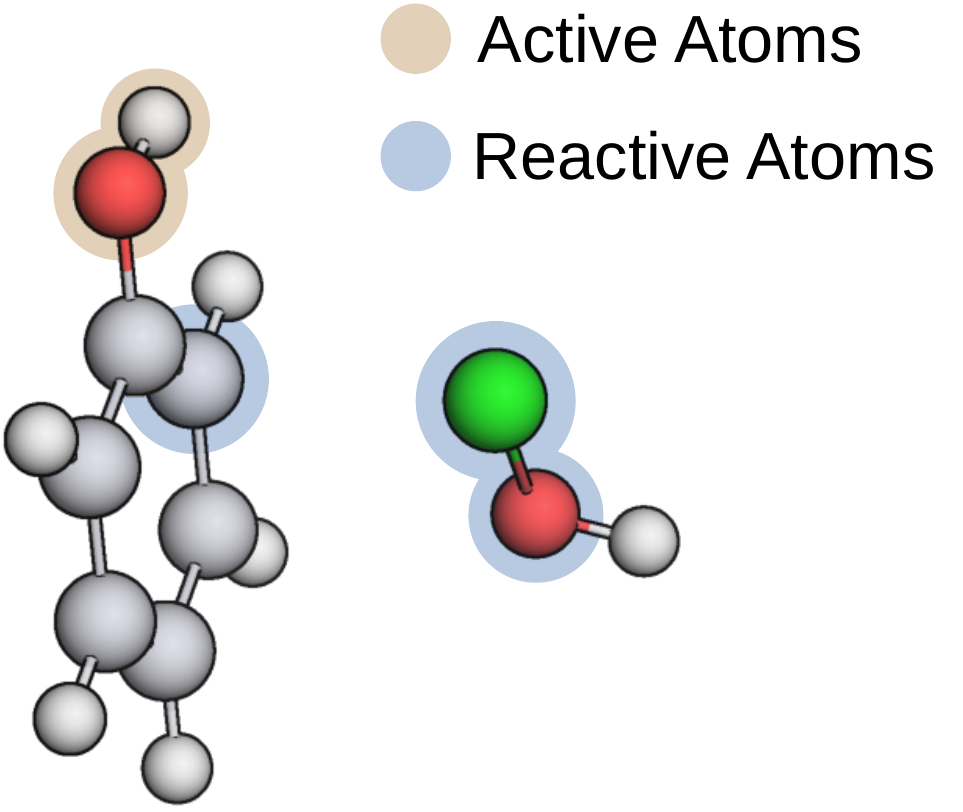}
  \caption{A reactive complex assembled of one phenol and one hypochlorous acid molecule.
    Carbon atoms are depicted in gray, hydrogen atoms in white, oxygen atoms in red, and the chlorine atom in green.
    Reactive atoms are highlighted in blue, active atoms in gold.
  }
  \label{fig:atomTypes}
\end{figure}

In the automated reaction mechanism exploration algorithms implemented in our Chemoton software,\cite{Simm2017,Unsleber2022a}
\textit{reactive atoms} restrict the reaction coordinates of bond forming or breaking events which will be screened for suitable transition state guess structures, as described in Ref.~\citenum{Unsleber2022a}.
In this work, we refer to atoms of reactants that likely interact with solvent molecules (e.g., through hydrogen bonds) as \textit{active atoms}.
The \textsc{Kingfisher} algorithm then automatically constructs quantum (QM) regions around the combined set of all \textit{reactive} and \textit{active} atoms;
an example of a reactive complex of two reactants with these atom types is shown in Figure~\ref{fig:atomTypes}.

\subsection{The \textsc{Kingfisher} Pipeline}\label{sec:structurePrep}

\begin{figure}[H]
  \centering
  \includegraphics[width=1.0\textwidth]{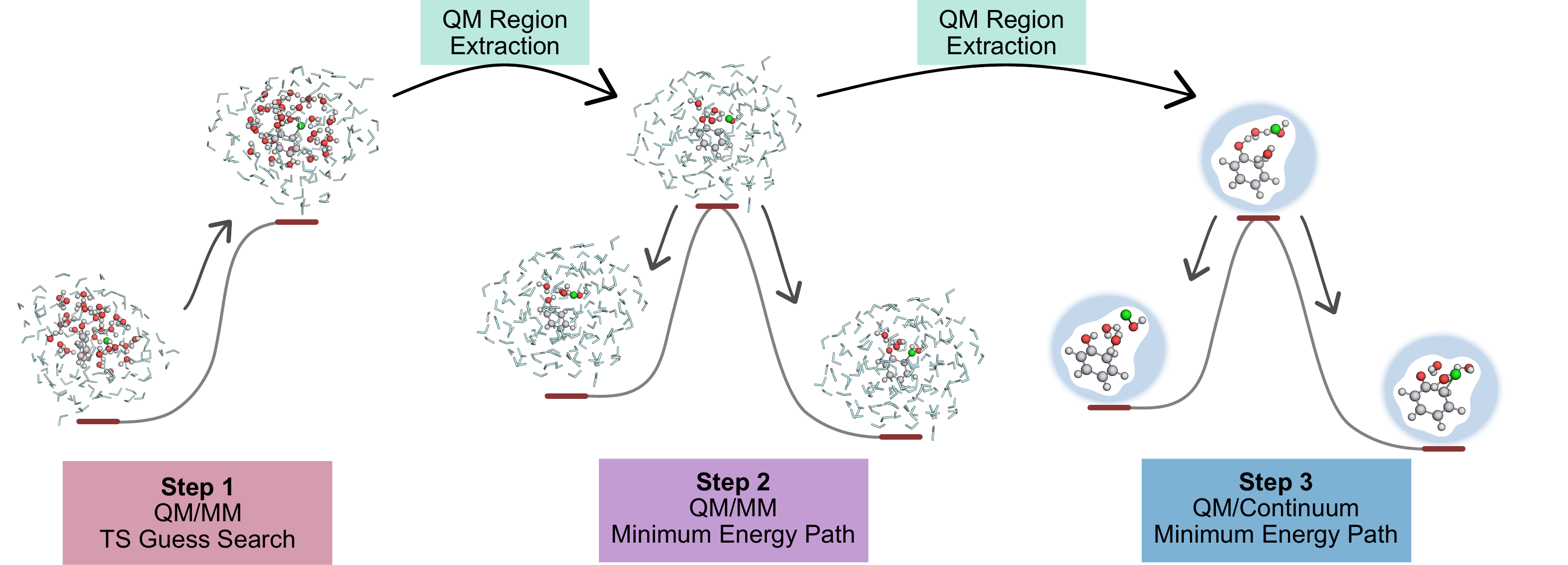}
  \caption{Schematic overview of the three steps of the \textsc{Kingfisher} pipeline.
    Structures in Step 1 are a partially optimized reactive complex and a TS guess.
    Step 2 and Step 3 contain optimized stationary points along the minimum energy paths.
    Atoms modeled by a QM model are represented with spheres, water molecules modeled by a MM model are represented as light blue sticks.
    Carbon atoms are depicted in gray, hydrogen atoms in white, oxygen atoms in red, and the chlorine atom in green.
  }
  \label{fig:KingfisherOverview}
\end{figure}

The intention of \textsc{Kingfisher} is to produce a cluster-continuum MEP for the smallest solvent-molecule cluster necessary to describe the active involvement of solvent molecules in the reaction. This information is deduced from a large system-focused QM/MM hybrid 
model of the individual solvation situation, which conceptually requires three steps (see Figure~\ref{fig:KingfisherOverview} for a graphical representation):

\begin{enumerate}
  \item Extraction of a transition state guess structure from a lQM/MM model with a large QM (lQM) region.
        The larger the lQM region is, the smaller we can expect any bias with respect to number and position of active solvent molecules to be.
        Different orientations of solvent molecules are sampled. In contrast to our previous work\cite{Csizi2024}, where we applied double-ended QM/MM elementary step searches,
  single-ended elementary step searches are then launched within this lQM/MM model.

  \item Next, \textsc{Kingfisher} obtains a MEP for this transition state guess, but with
  a QM/MM model where the QM region is medium-sized (mQM). This reduction im QM region size reduces the cost for the computationally more demanding tasks in this step.
  These tasks involve a mQM/MM transition state optimization followed by an intrinsic reaction coordinate calculation with subsequent optimization of the end points.

  \item Finally, analysis of the mQM/MM transition state obtained and extraction of a minimal number of solvent molecules is conducted by \textsc{Kingfisher}. The solvent molecule participation in the reaction is measured by their involvement in the lowest normal mode, which describes the collective motion of the nuclei along the reaction coordinate.
  It is, as usual, obtained as the eigenvector of the lowest (negative) eigenvalue of the Hessian matrix at the TS structure. The analysis of this normal mode yields the smallest microsolvated QM region (sQM), because it includes all solvent molecules that contribute directly to the reaction coordinate.
  For this sQM region combined with dielectric embedding, \textsc{Kingfisher} then obtains a MEP in a QM-only calculation, which facilitates the use of electronic structure methods that can be computationally more demanding and hence more accurate than those employed in the lQM/MM and mQM/modelling steps.
This last step minimizes the requirement\cite{Bensberg2022} of sampling many configurations to capture simple stabilizing effects on the relevant solute-solvent cluster and it eliminates the challenge regarding the fidelity of the description of interactions between the two models of a hybrid model.
\end{enumerate}

In the following subsections, we describe technical aspects within these three steps in detail.

\subsection{System-focused Selection of the QM Region}\label{sec:QMregionMMPara}
  
For the QM/MM hybrid model of solvation, we exploit our system-focused atomistic model (SFAM) expressed as a quantum-chemically parametrized force field that is derived specifically for the system at hand.\cite{Brunken2020,Brunken2021a}
Compared to our previous work,\cite{Brunken2021}
the QM region is not cut out of a large covalently bound system, but forms a solute-solvent cluster where only weak interactions among solvent molecules are to be dissected.
Hence, the methodology presented in Ref.~\citenum{Brunken2021a} can be simplified as follows.
The QM-region construction relies on spheres around the set of \textit{reactive} and \textit{active atoms} (denoted as \textit{ra-atoms}) of the reactants (compare Fig.~\ref{fig:atomTypes}).
All atoms of the reactive complex are always part of the QM region.
If any atom of a solvent molecule is within the sphere around an \textit{ra-atom}, all of the atoms of the solvent molecule will be included in the QM region.

The radius of the sphere around every \textit{ra-atom}, $r_\text{s}$, that spans part of the QM region can be chosen to be a fixed radius, $r_c$, for all atoms.
However, a single fixed radius might not be sufficient for reactants with \textit{ra-atoms} that are very different in size, as, for instance, oxygen and iodine.
Alternatively, the radius of the sphere around all \textit{ra-atom} can be made dependent on the type of atom, hence having a type-depended sphere radius for each \textit{ra-atom},
\begin{equation}
  \label{eq:adaptiveSphereRadius}
  r_s =
  \begin{dcases}
    r_c, \\
    s~(r_\text{cov,ra} + r_\text{cov,H})
  \end{dcases}
\end{equation}
where $r_\text{cov,ra}$ and $r_\text{cov,H}$ are the covalent radii of the \textit{ra-atom} and a hydrogen atom, respectively,\cite{Rumble2019} and $s$ is a scaling factor.
This definition allows us to include hydrogen bonded solvent molecules in protic solvents more easily, where active involvement of solvent molecules is most prominent.
$r_\text{cov,H}$ can, in principle, be exchanged for any solvent-specific radius; for instance, it can be replaced by $r_\text{cov,Cl}$ if the solvent is dichloromethane.

To determine which atoms are within the sphere of a \textit{ra-atom}, the Euclidean distances between the \textit{ra-atom} to all other atoms are calculated.
If the distance $d_{i}$ between the \textit{ra-atom} with Cartesian coordinates $\textbf{r}_\text{ra}$ and an atom $i$ with Cartesian coordinates $\textbf{r}_\text{i}$,
\begin{equation}
  d_{i} = \|\mathbf{r}_i - \mathbf{r}_{\text{ra}}\|,
\end{equation}
is smaller than the sphere radius $r_s$ of this \textit{ra-atom}, it will be considered to be part of the QM region $R_j$ around the \textit{ra-atom} $j$,
\begin{equation}
  R_j = \{i \mid d_i \le r_s\}
\end{equation}
The algorithm is applied for all \textit{ra-atoms}, resulting in a total set of atoms $R_\text{tot}$ lying within the spheres of all \textit{ra-atoms}.
\begin{equation}
  R_{\text{tot}} = \bigcup_j R_j.
\end{equation}

All molecules not included in the QM region belong to the MM region and will be described with a classical force field.
For a generally applicable, black-box approach of \textsc{Kingfisher}, where arbitrary reactant and solvent molecules can be involved, we would require a very general force-field parametrization that is not available. Therefore, and for the sake of maintaining accuracy and consistency, we employ the system-specific on-the-fly parametrization of SFAM\cite{Brunken2020} for all molecules in the MM region.
The reactants and solvent molecules are parametrized before solvation of the reactants is studied and the obtained parameters can be re-used for the solvent (or, if already present in a database, read from file).
Dependent on the reference method (compare Section~\ref{sec:compMethod}) and the size of the solvent and reactant molecules, the generation of the SFAM parameters of each molecule takes between a few seconds and a couple of minutes.
Here, the bottleneck for each molecule is the calculation of the Hessian, from which the parameters are then derived.
For details on the generation we refer the interested reader to Ref.~\citenum{Brunken2020}.
We note that our system-specific SFAM force-field may also be replaced by a general machine learning potential\cite{Batatia2022,Chen2022}
or in the framework of a lifelong machine learning potential\cite{Eckhoff2023},
when their general applicability has been fully established.

For each step outlined in \ref{sec:structurePrep}, the QM region is re-defined by decreasing $r_s$ systematically.
This ensures feasibility for the computationally more demanding tasks in step 2 of the pipeline and eventually results in the definition of the minimum solute-solvent cluster model in step 3.

\subsection{QM/MM Transition-State Guess Extraction}\label{sec:QMMMTSGuess}

The first step toward a QM/MM model is the construction of an initial solute-solvent complex consisting of the reactive complex (RC) and solvent molecules. 
With the \textit{reactive} and \textit{active} atoms (see Fig.~\ref{fig:atomTypes}), solvent molecules,
and their SFAM parameters at hand, our pipeline assembles a RC\cite{Simm2017, Unsleber2022a} and randomly places two shells of solvent molecules around it, as described in Ref.~\citenum{Simm2020}.
For example, in the case of formaldehyde in water, the two shells consist of about 100 water molecules.
Other approaches deduce the placement of solvent molecules based on the free solvation energy derived from molecular dynamics simulations.\cite{Steiner2021,Talmazan2023}

For one RC, different placements of solvent molecules can be obtained with a seed for a random number generator, which results in a specific number of reaction trials.
We then optimize the position of all solvent molecules described in the MM region in terms of their MM energy, keeping the atoms of the RC frozen.
As a result, a MM optimized microdroplet is obtained.
The initial QM region is then constructed with spheres around the \textit{ra-atoms} with a fixed radius $r_c$ of \SI{4.5}{\angstrom}.
The positions of the large QM region (lQM) are then optimized in terms of the lQM/MM energy with ten times larger convergence criteria than in subsequent structure optimizations.
In this optimization, the full system is described by the lQM/MM hybrid model while the atoms of the RC are still kept frozen.
In principle, the optimization procedure can be iterated, because after each optimization step the lQM is redefined by the spheres of fixed radii around the \textit{ra-atoms}.
By default, the optimization is run twice, resulting in optimized solvent molecules in the lQM encapsulated by the solvent molecules described by the MM model.

\begin{figure}[!htb]
  \centering
  \includegraphics[width=0.95\textwidth]{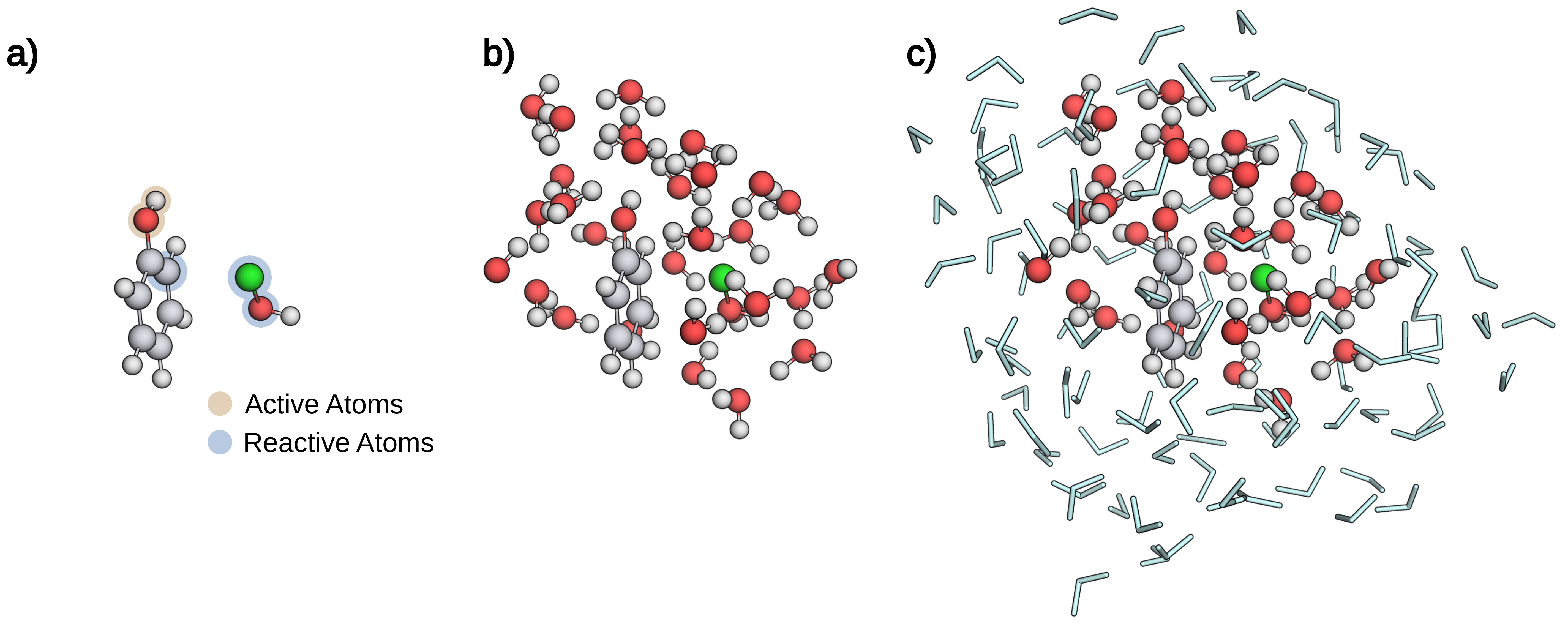}
  \caption{
    \textbf{a)} Reactive complex \textbf{b)} Atoms of the lQM region. \textbf{c)} Full lQM/MM system before the elementary step search.
    Atoms within the QM region are represented as spheres, water molecules modeled by a MM model are represented as light blue sticks.
    Carbon atoms are depicted in gray, hydrogen atoms in white, oxygen atoms in red, and the chlorine atom in green.
  }
  \label{fig:lQMMM}
\end{figure}

\textsc{Kingfisher} then searches for a TS structure guess by running our Newton Trajectory 2 (NT2) algorithm\cite{Unsleber2022a} with the full lQM/MM system.
If the NT2 algorithm extracts a TS guess structure, \textsc{Kingfisher} calculates the partial Hessian of the lQM region (keeping the solvent molecules of the MM region frozen), as introduced in Ref.~\citenum{Csizi2024}.

\subsection{Microsolvated QM/MM Minimum Energy Paths}\label{sec:QMMMMEP}

\begin{figure}[ht]
  \centering
  \includegraphics[width=0.95\textwidth]{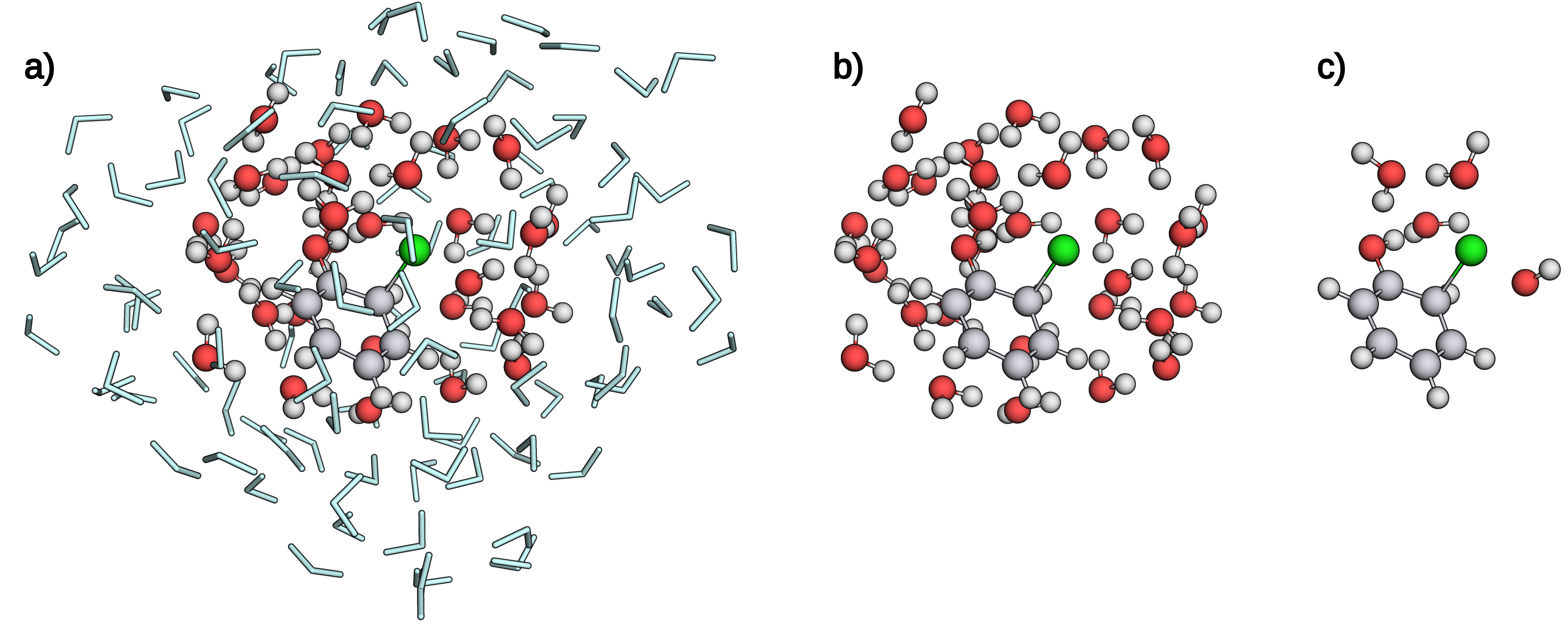}
  \caption{
    \textbf{a)} Transition state guess of the full lQM/MM system. \textbf{b)} Transition state guess of the lQM region. \textbf{c)} Extracted mQM region with three water molecules.
    Atoms within the QM region are represented as spheres, water molecules modeled by a MM model are represented as light blue sticks
    Carbon atoms are depicted in gray, hydrogen atoms in white, oxygen atoms in red, and the chlorine atom in green.}
  \label{fig:lQM2mQM}
\end{figure}

Due to the size of the lQM region and the computational cost of partial Hessian calculations, the number of atoms in the QM region must be reduced in order for efficient routine QM/MM transition state optimizations.
We achieve this pruning by an algorithm which analyzes the normal mode that describes the reaction coordinate best.
Here, the normal mode is the eigenvector of a non-mass weighted Hessian, describing distortions of all atoms in Cartesian coordinates.
First, this eigenvector of the TS guess structure (or of the TS structure in the subsequent section) must be selected, and second, analyzed to extract the active solvent molecules.

Given $N$ atoms, an eigenvector (normal mode) $\mathbf{n} \in \mathbb{R}^{N \times 3}$ of the lQM partial Hessian with high contributions of the \textit{reactive atoms} and low
eigenvalue is selected.
The algorithm for selecting the eigenvector is introduced here.
The TS optimization algorithm employed in this work operates with the non-mass weighted Hessian, and therefore, the described algorithm utilizes the non-mass weighted Hessian.
Every eigenvector $\textbf{n}_i$ obtained from this non-mass weighted Hessian with a negative eigenvalue $w_i$ is analyzed.
First, the $i$th vector $\mathbf{n}$ is converted to mass-weighted coordinates by elementwise multiplication,
\begin{equation}
  \label{eq:massWeighting}
  \mathbf{n}^{(m)}=\mathbf{n} \odot \sqrt{\mathbf{m}},
\end{equation}
where the vector $\mathbf{m}$ contains the masses of the atoms, $\mathbf{m} = [m_1, m_2, \ldots, m_N]$.
The eigenvector is mass-weighted to damp the otherwise large contributions of light atoms, such as hydrogen atoms, in relation to heavy atoms.
The resulting $\mathbf{n}^{(m)}$ vector is normalized according to
\begin{equation}
  \label{eq:normalizaton}
  \mathbf{n}^{(m)}_{\text{norm}} = \frac{\mathbf{n}^{(m)}}{\|\mathbf{n}^{(m)}\|}.
\end{equation}
The $i$th total contribution $C$ of the \textit{reactive atoms} in the $i$th vector $\mathbf{n}^{(m)}_{\text{norm}}$ is then calculated according to
\begin{equation}
  C_i = \sum_{j \in \text{reactive atoms}}
       {{n}^{(m)}_{\text{norm},j,x}}^2 + {{n}^{(m)}_{\text{norm},j,y}}^2 + {{n}^{(m)}_{\text{norm},j,z}}^2.
\end{equation}
Concerning the negative eigenvalues, the minimum eigenvalue $w_{\text{min}} = \min_{i \in \textit{w}} w_i$ is selected to normalize the eigenvalues with $w_{\text{norm},i} = w_i / w_\text{min}$.
Both, $w_{\text{norm},i}$ and $C_i$ of an analyzed eigenvector are in the interval $[0,1]$.
With these two properties, the selection of an eigenvector is based on the score $s_i$ of an eigenvector.
The score $s_i$ is obtained by a weighted sum of $w_{\text{norm},i}$ and $C_i$
\begin{equation}
  s_i = 0.5~w_i + 0.5~C_i,
\end{equation}
where the weights for this study are fixed to be $0.5$.
The eigenvector with the highest score $s_i$ is selected and analyzed for the identification of the active solvent molecules.

To determine the active solvent molecules according to the selected eigenvector $\mathbf{s}$, $\mathbf{s}$ is processed according to Eq.~\eqref{eq:massWeighting} and Eq.~\eqref{eq:normalizaton} to obtain the mass-weighted, normalized selected eigenvector $\mathbf{s}_\text{norm}^{(m)}$.
The relative contribution of any atom $i$, $c_i$, is then given by
\begin{equation}
  c_i ={{s}^{(m)}_{\text{norm},i,x}}^2 + {{s}^{(m)}_{\text{norm},i,y}}^2 + {{s}^{(m)}_{\text{norm},i,z}}^2.
\end{equation}
The threshold to determine which atoms are active depends on the minimum contribution of the \textit{reactive atoms}, $c_{\text{min}} = \min_{i \in \textit{reactive\ atoms}} c_i$.
Atoms with higher or equal contributions compared to $c_{\text{min}}$ form the set of $\textit{involved\ atoms} = \{ i \mid c_i \geq c_{\text{min}}\}$.
The molecules of which the \textit{involved atoms} are part of are included in the mQM region.
The \textit{involved atoms} are combined with the \textit{active atoms} to form a set of \textit{relevant atoms}.
Then, the mQM region is constructed by employing spheres with atom-dependent radii, around the \textit{mQM relevant atoms} with a scaling factor $s$ of $2.6$ (cf. Eq.~\eqref{eq:adaptiveSphereRadius}).

After this re-definition of the QM region (transgressing from lQM to mQM), the MM degrees of freedom have to be relaxed while keeping the atoms of the mQM region fixed.
The partial Hessian of the TS guess structure is then re-calculated with the mQM/MM model.
For the subsequent mQM/MM TS optimization, the optimizer\cite{Bofill1994} follows the eigenvector of the mQM/MM partial Hessian with the lowest eigenvalue and the highest contributions of the \textit{reactive atoms}.
If the mQM/MM TS optimization is successful, the full mQM/MM MEP is generated through a mQM/MM intrinsic reaction coordinate
scan followed by a mQM/MM optimization of the endpoints of the scan.
The endpoints correspond to non-covalently bound solute-solvent clusters.
These are further disassembled into the separate covalently bound molecules of which the cluster consists.
These individual molecules are then separately optimized with the electronic structure model employed for describing the mQM region.

\subsection{QM Minimum Energy Paths}\label{sec:QMMEP}

Due to the large radii $r_s$ employed for the construction of the mQM region,
the TS of the mQM/MM MEP might still involve solvents which are not directly involved in the relevant eigenvector.
To minimize the number of possible configurations and avoid expensive sampling approaches,\cite{Bensberg2022} all residual solvent effects are modeled with an averaging continuum solvation model.
Therefore, the number of solvent molecules is further reduced.
For this, \textsc{Kingfisher} applies the same algorithm as described in Section~\ref{sec:QMMMMEP} and analyzes the eigenvector of the lowest eigenvalue obtained from the partial Hessian of the mQM/MM TS to identify atoms of solvent molecules as the \textit{involved atoms} if they have higher or equal contributions compared to the \textit{reactive atoms}.
Together with the \textit{active atoms}, they form the set of \textit{sQM relevant atoms}.

The emerging final small QM region (sQM) is constructed by employing spheres with atom-dependent radii around the \textit{sQM relevant atoms} with a scaling factor $s$ of $2.0$.

\begin{figure}[h]
  \centering
  \includegraphics[width=0.6\textwidth]{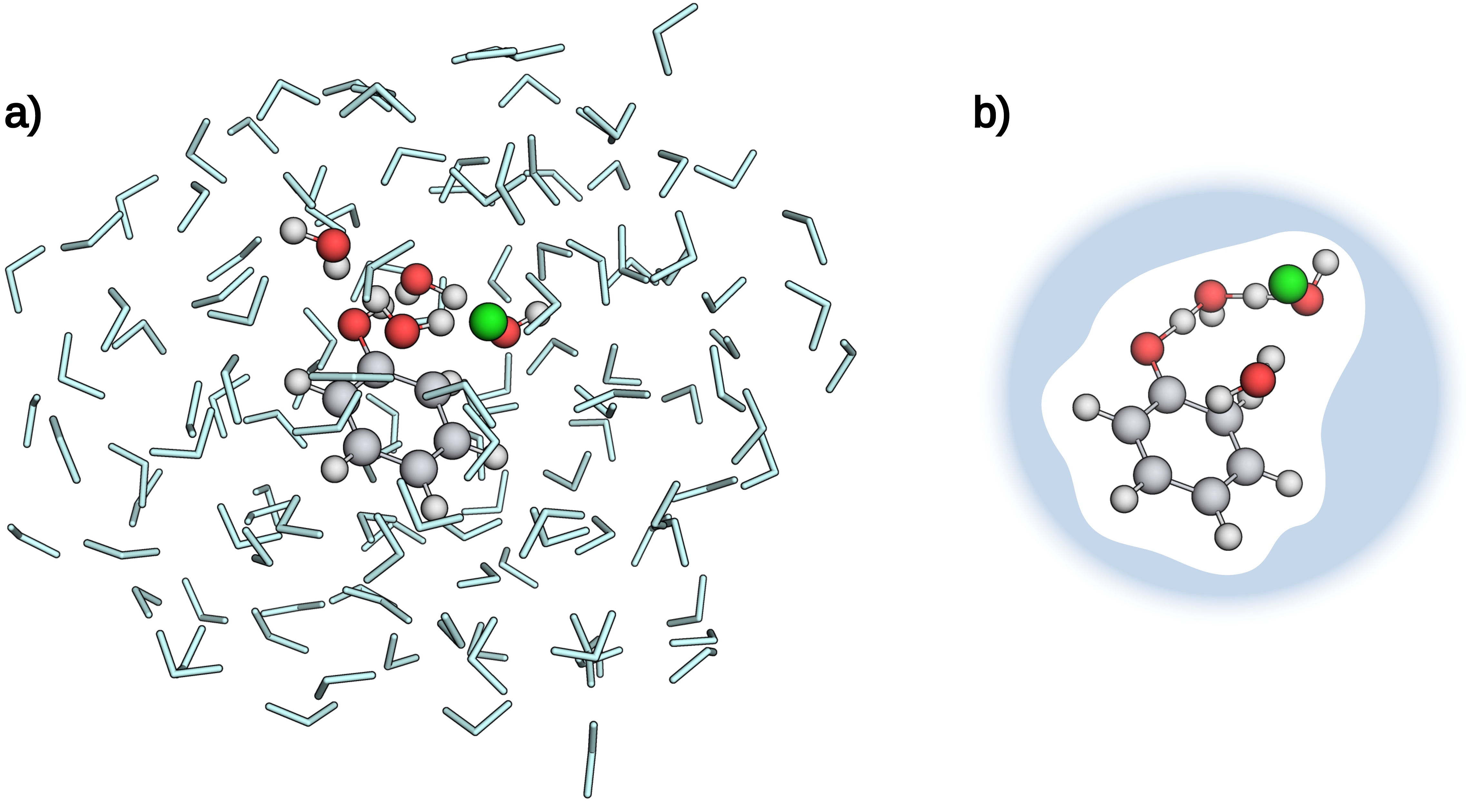}
  \caption{
    \textbf{a)} Optimized TS of the full mQM/MM system with three water molecules in the mQM region.
    \textbf{s)} Optimized TS of the sQM region with two water molecules in the sQM region, schematically embedded in the cavity constructed by the continuum solvation model.
    Atoms within the QM region are represented as spheres, water molecules modeled by a MM model are represented as light blue sticks
    Carbon atoms are depicted in gray, hydrogen atoms in white, oxygen atoms in red, and the chlorine atom in green.}
  \label{fig:mQM2sQM}
\end{figure}

This final sQM region is now solely described with a QM model.
To obtain a MEP with this final sQM region, a new elementary step search with \textsc{Chemoton} is launched,
describing the sQM with a QM model with a continuum solvation model (sQM/C), starting with a sQM/C TS optimization with the extracted sQM region as TS guess.
If the optimization is successful, the full sQM/C MEP is determined as described in Section~\ref{sec:QMMMMEP}.

\subsection{Algorithmic Workflow}

\begin{figure}[H]
  \centering
  \includegraphics[width=1.0\textwidth]{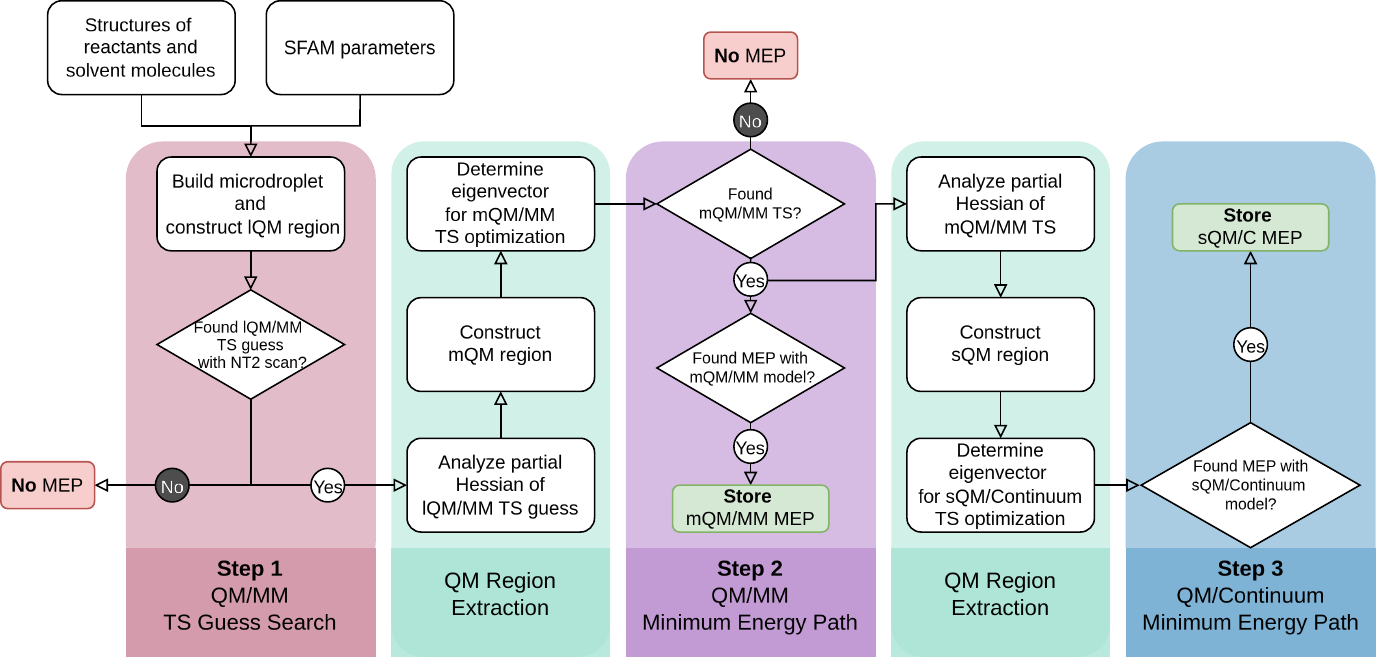}
  \caption{
    Flowchart of the individual steps of the \textsc{Kingfisher} model to obtain a QM/MM MEP and a QM/Continuum MEP.
  }
  \label{fig:KingfisherWorkflow}
\end{figure}

The complete workflow of \textsc{Kingfisher} is shown in Fig.~\ref{fig:KingfisherWorkflow}.
The coloring scheme follows that of the steps shown in Fig.~\ref{fig:KingfisherOverview} to clarify the technical requirements behind the overall concepts already introduced.

\section{Microsolvated Free Energy of Activation Model}\label{sec:FreeEnergyQM}

In the following, we introduce a model for the microsolvated free energy of activation (MiFEA) calculation in order to facilitate kinetic modeling and a direct comparison to experimental results.
We design the MiFEA model to be closely aligned to standard quantum chemistry free energy calculations so that it can be efficiently exploited for condensed phase reactions.
Naturally, such a simple model will be affected by errors, which, in turn, have a dramatic effect on reaction rate constants as the activation free energy enters those in the argument of an exponential.
However, the simplicity of the model may allow for a straightforward parametrized improvements (in a simple form or through an advanced $\Delta$-machine learning correction where reference data for some structures is extrapolated to other structures that lack this information).

Within Eyring's absolute rate theory,\cite{Eyring1935} a rate constant $k$ is obtained from the (Gibbs) free energy of activation $\Delta G^\ddag$ (in the ($N$,$p$,$T$)-ensemble for constant macroscopic particle number $N$, pressure $p$, and temperature $T$) according to
\begin{align}
	k = \frac{k_\text{B}T}{h}\exp\left(-\frac{\Delta G^\ddag}{RT}\right),\label{eq:Eyinring}
\end{align}
where $k_\text{B}$ is the Boltzmann constant, $T$ the absolute temperature, $h$ the Planck constant, and $R$ the molar gas constant.
The change in Gibbs free energy
\begin{align}
\label{eq:DeltaG}
	\Delta G^{\ddag} = G^\text{TS} - G^\text{R},
\end{align}
defines the activation free energy, where $G^\text{TS}$ and $G^\text{R}$ are molar Gibbs free energies of the TS structure and the minimum structure of a reactant R, respectively.

The Gibbs free energy $G$ of a thermodynamic state ($N$,$p$,$T$) is the difference between its enthalpy $H$ and its temperature-weighted entropy $S$,
\begin{align}
	G(T) &= H(T) - TS(T) \label{eq:G},
\end{align}
where we only explicitly denote the temperature dependence because the pressure dependence will be approximated from the constant-volume canonical ensemble expressions through the ideal gas law and all quantities will be considered for \SI{1}{mol} solute reactants.
The Gibbs free energy of activation then reads
\begin{align}
\label{eq:stdGact}
	\Delta G^\ddag &= \Delta H^{\ddagger} - T \Delta S^{\ddagger}.
\end{align}

The enthalpy is the sum of the internal energy $U$ and the pressure-volume work, $H=U+pV$.
In the standard quantum chemical model of gases, the internal energy $U$ and the entropy $S$ can be calculated from the (canonical) molecular partition function for decoupled degrees of freedom of translation, rotation, vibration, and electronic motion.
Since our clusters will, in general, exhibit no symmetry, we consider the microsolvated cluster as a non-symmetric rigid rotor of point group $\text{C}_1$ (corresponding to a symmetry number $\sigma$ of $1$).
Each of the $N$ clusters is considered to move freely in the constant macroscopic volume $V$ so that the quantum mechanical particle in the box model can be exploited for the description of its translation in gaseous state.
Without loss of generality, we also assume a spin multiplicity of one and a sufficient energetical separation of the electronic ground state from the first electronically excited stated so that no electronic degeneracy needs to be considered.
All vibrations are considered in harmonic approximation, which introduces a well-defined, albeit approximate model of decoupled molecular vibrations.
Then, the internal energy $U$ and the entropy $S$ are decomposed as
\begin{align}
	U(T) &= U_\text{el} + U_\text{tra}(T) + U_\text{rot}(T) + U_\text{vib}(T),\\
	S(T) &= S_\text{tra}(T) + S_\text{rot}(T) + S_\text{vib}(T),
\end{align}
where $U_\text{el}$ is the electronic energy $E_\text{el}$ obtained from some electronic structure model. $U_\text{tra}$ and $S_\text{tra}$ are the translational, $U_\text{rot}$ and $S_\text{rot}$ the rotational, and $U_\text{vib}$ and $S_\text{vib}$ are the vibrational contributions.

The translational contribution for the particle-in-a-box model yields\cite{Swendsen2019}
\begin{align}
	U_\text{tra} &= \frac{3}{2}RT \label{eq:Utrs}\\
	S_\text{tra} &= R \left(
		\frac{5}{2} +
		\ln\left(
			\left(\frac{2\pi m k_\text{B}T}{h^2}\right)^\frac{3}{2}
			\frac{k_{B}T}{p}
		\right)
	\right) \label{eq:Strs}
\end{align}
with $m$ being the total mass of the cluster structure.
With the rigid-rotor partition function, the rotational contributions read
\begin{align}
	U_\text{rot} &= \frac{3}{2}RT \label{eq:Urot}\\
	S_\text{rot} &= R \left(
		\frac{3}{2} +
			\frac{3}{2}\ln\left(
				\frac{8\pi^{\frac{7}{3}}k_\text{B}T}{h^2}
				I_{a} I_{b} I_{c}
			\right)
		\right) \label{eq:Srot}
\end{align}
with the principal moments of inertia, $I_{a}$,  $I_{b}$, and $I_{c}$ and of the cluster structure under consideration.
Within the harmonic approximation that is easy to evaluate by diagonalizing the mass-weighted Hessian matrix,\cite{Neugebauer2002}
the harmonic frequency $\nu_i$ are obtained, and hence, the internal vibrational energy and vibrational entropy,
\begin{align} 
    U_\text{vib} &= N_\text{A}h\sum_i
						 \nu_i\left[
						 	\frac{1}{2} + \left(\exp\left(\frac{h\nu_i}{k_\text{B}T}\right) - 1\right)^{-1}
						 \right]\\
	S_\text{vib} &= R \sum_i
		\frac{h\nu_i}{k_\text{B}T}
		\left(\exp\left(\frac{h\nu_i}{k_\text{B}T}\right)-1\right)^{-1} -
		\ln\left(1 - \exp\left(-\frac{h\nu_i}{k_\text{B}T}\right)\right).
	\label{eq:Svib}
\end{align}
Note that we have omitted an index for the specific cluster structure under consideration for the sake of simplicity.

As is obvious from the vibrational entropy expression in Eq.~\eqref{eq:Svib}, low-lying vibrational frequencies $\nu_i$ contribute the most to the entropy $S_\text{vib}$ and their shifting during a reaction will dominate the entropy change. However, the harmonic approximation will tend to break down for these frequencies as the associated motions cannot be well represented by a parabolic potential energy surface.
It is, however, not straightforward to correct for this deficiency without creating a major computational effort.
Simple cures are typically associated with specific uncertainties.
For instance, it has been proposed to improve the harmonic approximation by invoking an explicit anharmonic potential along a normal coordinate,\cite{Piccini2014}
where the vibrational coordinate still remains in the harmonic approximation and the vibrational partition function is no longer obtained in closed form.
This has also been extended to determine Gibbs free energies during adsorption processes.\cite{Rybicki2022}
It has also been proposed to consider low-frequency vibrations as hindered rotations in a one-dimensional rotor model with a moment of inertia that is associated with frequency $\nu_i$.\cite{Grimme2012,Conquest2021}

The enthalpy of activation reads
\begin{align}
	\Delta H^{\ddag} = \Delta U^{\ddag} + p\Delta V^{\ddag}.
\end{align}
Assuming a negligible change in volume, $\Delta V^{\ddagger} = 0$, $\Delta H^{\ddagger}$ only depends on the internal energy of activation $\Delta U^{\dagger}$.
Considering Eq.~\eqref{eq:Utrs} and Eq.~\eqref{eq:Urot}, the translational and rotational internal energies of the TS and the minimum structure R cancel out and only the electronic contribution, $\Delta E^\ddagger_\text{el}$, and the vibrational part, $\Delta U^\ddagger_\text{vib}$, survive in the standard model,
\begin{align}
	\Delta H^{\ddagger} = \Delta E^\ddagger_\text{el} + \Delta U^\ddagger_\text{vib}. \label{eq:Hact}
\end{align}
For the entropy of activation $\Delta S^\ddagger$, only the translational entropies of the two states cancel each other, the rotational and vibrational contributions survive,
\begin{align}
\Delta S^{\ddagger} = \Delta S^\ddagger_\text{rot} + \Delta S^\ddagger_\text{vib}. \label{eq:Sact}
\end{align}
If we consider the structural changes from the reactant R to the TS to be minimal,
the two sets of principal moments of inertia will be similar and the difference in rotational entropy small, compared to change in vibrational entropy.

The assumptions of free-particle translation and free rotations of a cluster are no suitable description for the condensed phase, and hence, further corrections will be required for reactions in solution.
The largest contribution to the enthalpy of activation in Eq.~\eqref{eq:Hact} will usually originate from the change in electronic energy between R and TS, provided that the barrier is not vanishing.
Stabilizing electrostatic solvation effects will be considered with a polarizable continuum model in the electronic structure calculation, which we consider to be efficient for our MiFEA model.

As our MEPs are represented on a single potential energy surface (PES) and always start from an optimized solute-solvent cluster, the change in entropy from the solute in solution to the TS in solution will be underestimated because
these clusters are known to form more stable hydrogen bonds than observed in experiment.\cite{Bensberg2022}
To capture the change in entropy for the non-associated solute to the solute-solvent cluster along the MEP, molecular dynamics simulations would be required to sample the relevant part of configuration space.
However, an advanced sampling approach, although far more reliable, is computationally very demanding\cite{Giesen1994,Ratkova2015,Besora2018,Garza2019,Gorges2022} and cannot be applied in routine high-throughput studies of vast chemical reaction networks.
Hence, we yet lack this entropy penalty of forming the solute-solvent cluster in our MiFEA model and expect the resulting free energies of activation to underestimate experimental results, especially the entropy of activation.

\begin{figure}[H]
  \centering
  \includegraphics[width=0.65\textwidth]{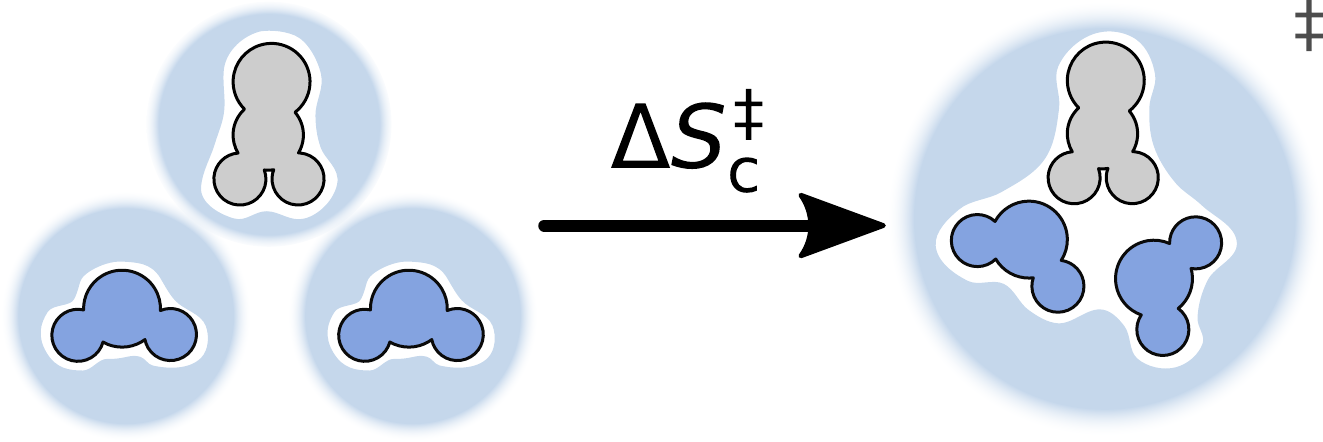}
  \caption{
    Schema to determine the change in cavity entropy from one solute $\text{R}'$ in gray and two explicit solvent molecules L, in dark blue, embedded in their cavities of the continuum, in light blue, to the TS with its own cavity in the continuum.
  }
  \label{fig:cavityScheme}
\end{figure}

Since, in \textsc{Kingfisher}, a reactant R is a solute-solvent cluster consisting of the isolated solute structure $\text{R}'$ and a number of explicit solvent molecules embedded in a dielectric continuum,
we may consider this minimum structure R to be assembled from $\text{R}'$ and solvent molecules from the continuum.
This assembly corresponds to a loss in entropy.
Hence, we introduce the cavity entropy to correct to some degree the lacking entropy penalty, schematically depicted in Fig.~\ref{fig:cavityScheme}.

We derive the change in cavity entropy of activation $\Delta S_{\text{c}}^{\ddag}$ from the isolated solute structure $\text{R}'$ and the sum of $n$ isolated solvent molecules L in the continuum to the TS,
\begin{align}
	\Delta S^{\ddagger}_\text{c} = S^\text{TS}_\text{c} - (S^\text{{R}'}_\text{c} + n S^\text{L}_\text{c}).\label{eq:Scact}
\end{align}
The cavity entropies $S_{\text{c}}$ of the isolated solute $\text{R}'$ and solvent L and the TS can be calculated with an approach proposed in Ref.~\citenum{Garza2019}, which relies on the relative permittivity $\epsilon_\text{r}$ and the volumes of the solute molecule and the solvent molecule.
The cavity entropy is, in general, defined as
\begin{equation}
	S_{\text{c}}^{\epsilon} = \frac{G_{\text{c}}}{T},
\end{equation}
where the free energy of cavity formation $G_{\text{c}}$ can be defined in polarizable continuum models\cite{Pierotti1976,Tomasi2005,Garza2019} with
\begin{align}
	\label{eq:gc}
	G_{\text{c}} &= -\ln(1 - y) +
						R_\text{s}\frac{3y}{1 - y} + R^2_\text{s}\left(\frac{3y}{1 - y} +
						\frac{9}{2}\left(\frac{y}{1 - y}\right)^2\right) \\
	y &= \frac{3}{4\pi}\frac{\epsilon_r - 1}{\epsilon_r + 2}.
\end{align}
$R_\text{s}$ in Eq.~\eqref{eq:gc} is the ratio of the scaled radii derived from the volume $V_\text{solute}$ and $V_\text{solvent}$ with
\begin{equation}
	\label{eq:radiiRatio}
	R_\text{s} = \left(\frac{V_{\text{solute}}}{V_{\text{solvent}}}\right)^{\frac{1}{3}}.
\end{equation}
Since $R_\text{s}$ is a ratio, Eq.~\eqref{eq:radiiRatio} can be expected to be accurate as long as the volumes $V_\text{i}$ are determined in the same way.
Adding $\Delta S^{\ddagger}_\text{c}$ from Eq.~\eqref{eq:Scact} to Eq.~\eqref{eq:Sact} and with Eq.~\eqref{eq:stdGact} we derive a Gibbs free energy of activation accounting for the entropy change due to the cavity formation with
\begin{align}
\label{eq:cavGact}
\Delta G^\ddag_\text{c} &= \Delta G^\ddag - T \Delta S^{\ddagger}_\text{c}.
\end{align}
With this last brick in our MiFEA model, we can compare the electronic energy of activation, $\Delta E^{\ddag}_\text{el}$ with the Gibbs free energies of activation, $\Delta G^\ddag$ and $\Delta G^\ddag_\text{c}$ in order to illustrate the different effects in free energy calculations, especially for entropy considerations.

\section{Computational Methodology}\label{sec:compMethod}

We implemented \textsc{Kingfisher} into our \textsc{SCINE} software framework\cite{Weymuth2024}. It is accessible through 'jobs' in the \textsc{SCINE Puffin} module\cite{QcscinePuffinAll2024} which orchestrates quantum chemical calculations.
Steps 1 and 2 are executed by the \texttt{scine\_kingfisher} job, step 3 by the \texttt{scine\_react\_ts\_guess} job.
All reaction exploration trials
were managed and analyzed with our automated exploration module \textsc{SCINE Chemoton}\cite{Unsleber2022a,QcscineChemotonAll2024}.
Molecular graph representations of the individual structures were obtained with \textsc{Molassembler}.\cite{Sobez2020,QcscineMolassemblerAll2024}
All calculations were executed on a high performance computing infrastructure with a \textsc{SCINE Puffin} singularity container and on local machines with a \textsc{SCINE Puffin} instance,
the execution being highly parallelized by running hundreds of \textsc{SCINE Puffin}s simultaneously on both architectures, enabling the high-throughput feature required to sample sufficiently many reaction trials.
All reaction trials
and results were stored in a \textsc{SCINE Database}.\cite{QcscineDatabaseAll2024}
All structure modifications -- such as structure optimizations, Newton trajectory scans, or intrinsic reaction coordinate optimizations --
have been carried out with \textsc{SCINE ReaDuct}.\cite{Vaucher2018,QcscineReaductAll2024}

Electronic structure calculations were executed with the \textsc{SCINE} Calculator\cite{QcscineCoreAll2024,QcscineXtbAll2024} interface which allows for separate energy, gradient, and Hessian calculations required by any of the structure manipulation steps.
In steps 1 and 2 of the model, the QM region was described with the semiempirical tight-binding method GFN2-xTB\cite{Bannwarth2019}, while the molecular mechanics calculations were based on the SFAM model\cite{Brunken2020} and executed with \textsc{Swoose}.\cite{QcscineSwooseAll2024}
For these GFN2-xTB/SFAM calculations, electrostatic embedding with SFAM point charges at the positions of the non-QM nuclei was employed.

The molecular volumes of the solute and the solvent, required in Eq.~\eqref{eq:radiiRatio}, were derived with molecular surfaces assuming van der Waals spheres for the atoms and the convex hull function implemented in the SciPy Spatial library.\cite{Barber1996}
The SFAM models for the various systems were parametrized with \textsc{Swoose}, employing the \textsc{Turbomole} program package (v7.4.1)\cite{Balasubramani2020} with the PBE exchange-correlation density functional\cite{Perdew1992, Perdew1996b} and semiclassical Becke-Johnson damped D3 dispersion corrections\cite{Grimme2010, Grimme2011} and a def2-SVP basis set\cite{Peterson2003b, Weigend2005a} to generate the reference data.

In step 3, electronic structure calculations were carried out with the \textsc{Orca} program package (v5.0.3)\cite{Neese2012, Neese2018, Neese2020} and the PBE functional\cite{Perdew1992, Perdew1996b} with semiclassical Becke-Johnson damped D3 dispersion correction\cite{Grimme2010, Grimme2011}, the def2-SVPD basis set\cite{Peterson2003b, Weigend2005a, Rappoport2010a}, and the conductor-like polarizable continuum solvation model.\cite{Barone1998, Garcia-Rates2020}

The thermochemical properties of step 3 were calculated assuming the rigid-rotor/harmonic-oscillator/particle-in-a-box model at a temperature of \SI{298}{K} and a pressure of \SI{1}{atm} assuming an ideal gas.

For each reaction investigated in the Results section, reaction pathways were identified automatically with the \textsc{Pathfinder}
algorithm implemented in \textsc{SCINE Chemoton},\cite{Turtscher2023} finding all paths with exactly one TS connecting reactants and products.

\section{Results and Discussion}\label{sec:results}

We now proceed to employ \textsc{Kingfisher} for three chemical reactions in solution.
The examples are chosen to illustrate the various features of \textsc{Kingfisher}.

We compare all \textsc{Kingfisher} results with results obtained by describing solvation solely through a dielectric continuum model, the conductor-like polarizable continuum solvation model implemented in the \textsc{Orca} program package,\cite{Barone1998, Garcia-Rates2020}
indicated by a dashed red line in the box plots below.
In case the reaction requires at least one solvent molecule, this molecule is added in this reference calculation.
Where available, we also compare our results to experimental observations reported in the literature.

\subsection{Reactions of Polarized Molecules}

The first group of reactants investigated possess polarized \textit{reactive atoms}.
Polarized, in this context, means that \textit{reactive atoms} have bonds with other atoms of significantly different electronegativity.
First, we applied our model to an example where the solvent is known to be one of the reactants.
Then, we probed a reaction where the solvent is not believed to be participating directly.

\subsubsection{Methanediol Formation from Formaldehyde}\label{sec:ch2o}

\begin{figure}[!htb]
  \centering
  \includegraphics[width=0.55\textwidth]{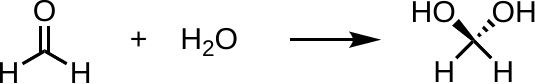}
  \caption{
    Methanediol formation from formaldehyde by reaction with water represented by Lewis structures.
  }
  \label{fig:kfpCH2O}
\end{figure}

\begin{figure}[H]
    \centering
    \includegraphics[width=0.55\textwidth]{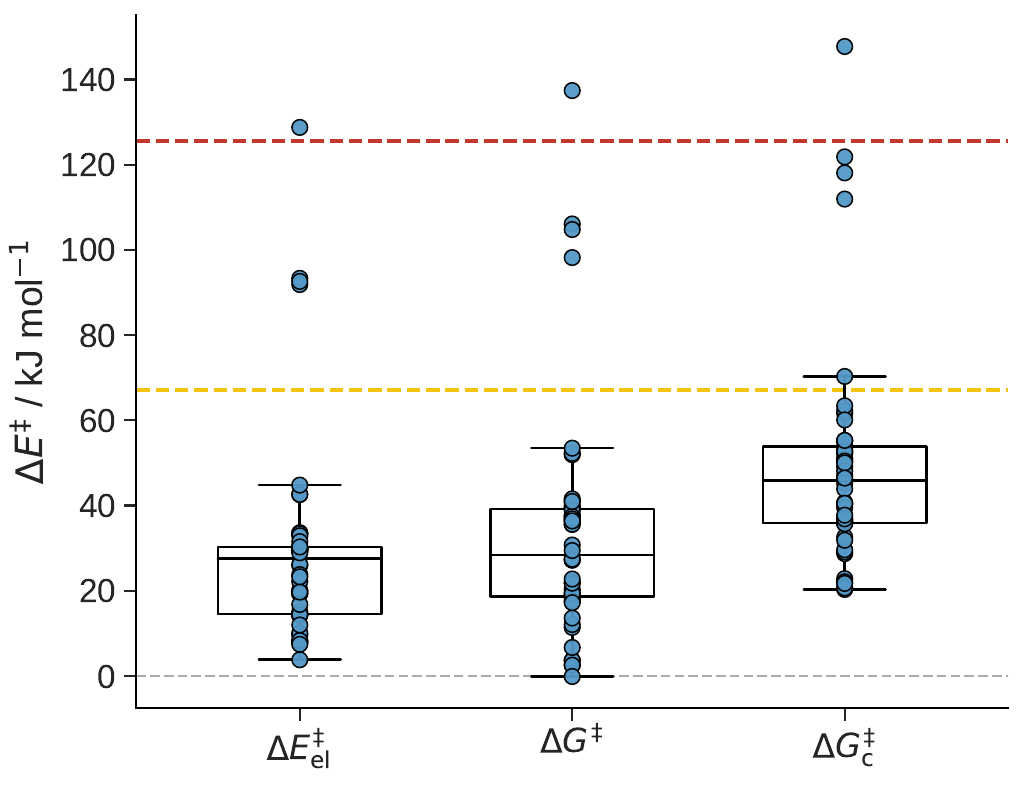}
    \caption{
      Comparison of sQM energies of activation of the reaction of formaldehyde with water to form methanediol.
      Left: electronic energy differences $\Delta E^{\ddag}_\mathrm{el}$;
      middle: standard free energy differences $\Delta G^\ddag$;
      right: free energy of activation corrected by a cavity entropy term $\Delta G^\ddag_\text{c}$.
      Box plots are added where the whiskers of the boxes indicate the minimum and maximum values.
      The box borders depict the second and third quartile, and the line in a box denotes the median of the energies of the group.
      The dashed red line indicates the free energy of activation $\Delta G^\ddag$ of \SI{125.6}{kJ~mol^{-1}} obtained by only considering continuum solvation.
      The yellow dashed line indicates the experimental reference value of \SI{67.1}{kJ~mol^{-1}}.\cite{Winkelman2002}
    }
    \label{fig:energyTypeCompCH2O}
\end{figure}

The exothermic and reversible formation of methanediol from formaldehyde in water\cite{Zavitsas1970} is one of the first reactions
where the effect of explicit solvent molecules on the reaction barrier has been studied computationally.\cite{Williams1983}
Subsequent theoretical studies investigated the reaction, considering two to five explicit water molecules.\cite{Wolfe1995,Zhang2008,Inaba2015,Wang2023}
With roughly evaluated 200 reaction trials starting from different initial placements of solvent molecules, \textsc{Kingfisher} automatically found 34 reactions within the mQM and 50 reactions within the sQM region.
We first investigate to what degree electronic activation energies and free energies of activation calculated according to the model discussed in section~\ref{sec:FreeEnergyQM}
reproduce the experimentally observed free energy of activation;
the different schemes are shown for the sQM model in Fig.~\ref{fig:energyTypeCompCH2O}.

Compared to the free energy of activation considering only continuum solvation
and the one water molecule as reactant,
most energies of activation shown in Figure~\ref{fig:energyTypeCompCH2O} are at least \SI{60}{kJ~mol^{-1}} lower and much closer to the experimental reference.
The four outliers in all three cases in Fig. ~\ref{fig:energyTypeCompCH2O} correspond to reactions where the TS is identical with the TS obtained from the continuum solvation model, i.e., considering only one water molecule.
Additional solvent molecules in these four TS structures form stabilizing hydrogen bonds and assist in the hydrogen atom transfer for the reacting water to the oxygen atom of the formaldehyde.

Compared to the experimental reference, $\Delta E^{\ddag}_\mathrm{el}$ and the free energy calculations with the standard model, $\Delta G^\ddag$ (see Eq.~\eqref{eq:stdGact}), both underestimate the experimental reference value on average by about \SI{40}{kJ~mol^{-1}}.
Winkelman and co-workers derived an experimental enthalpy of activation of \SI{21.8(27)}{kJ~mol^{-1}} and an experimental entropy of activation of \SI{-152(1)}{J~mol^{-1}~K^{-1}}.\cite{Winkelman2002}
Compared to our sQM $\Delta H^\ddag$ obtained from Eq.~\eqref{eq:Hact}, the mean enthalpy of activation of \SI{18(30)}{kJ~mol^{-1}} is in agreement with experiment.

However, the mean entropy of activation of \SI{-60(19)}{J~mol^{-1}~K^{-1}}, which depends only on the change in rotational and vibrational entropy (see Eq.~\eqref{eq:Sact}), underestimates the entropy penalty for the required arrangement of solute and solvent molecules in the sQM TS by about \SI{100}{J~mol^{-1}~K^{-1}}.
To describe this loss in entropy due to the formation of the solute-solvent cluster from the continuum, we introduced the cavity entropy $S_\text{c}^\epsilon$ in section~\ref{sec:FreeEnergyQM}.
With this correction term, the mean entropy of activation, $\Delta S^\ddag + \Delta S^\ddag_\text{c}$, is \SI{-107(24)}{J~mol^{-1}~K^{-1}}, capturing some of the experimentally observed loss in entropy.
With our MiFEA model, the resulting Gibbs free energy of activation $\Delta G^\ddag_\text{c}$ still underestimates the experimental one, but on average only by \SI{24}{kJ~mol^{-1}}.
Additional correction schemes for the entropy of vibration\cite{Grimme2012,Conquest2021}, as discussed in section~\ref{sec:FreeEnergyQM}, yielded only minor changes in the entropy of activation.
Hence, to keep our free energy calculations as simple as possible and as close to the standard model as possible, we are only considering the cavity entropy and analyze $\Delta G^\ddag_\text{c}$, as defined in Eq.~\eqref{eq:cavGact}.

We emphasize that a reliable calculation of the entropy contribution due to solvation is a well-known challenge (see, for instance, Refs.~ \citenum{Harvey2019, Conquest2021, Gorges2022, Tantillo2022, Cheong2024}).
As mentioned in section~\ref{sec:FreeEnergyQM}, to best capture the experimentally observed entropy, one would require to sample as many configurations as possible in a first-principles molecular dynamic simulation under periodic boundary conditions within a large unit cell in the canconical ensemble.
However, in the context of automated reaction exploration this is not a routinely applicable option, especially not in a high-throughput setting of automated reaction mechanism exploration.
In an exploration workflow, it is, however, no drawback to underestimate the energy of activation with a known error rather than to overestimate it (which could lead to a shut down of reaction channels because of artificially too high reaction barriers).
The latter might cause the resulting products to be excluded for further exploration steps, if kinetic screening is switched on.\cite{Bensberg2023,Bensberg2024}
A more promising alternative might be the application of a $\Delta$-machine learning approach, where accurate simulation data is obtained for some elementary steps in a reaction network and a Gaussian process is exploited to transfer this knowledge to similar structures (as demonstrated in Ref.~\citenum{Simm2018}).

\begin{figure}[H]
  \centering
  \includegraphics[width=0.95\textwidth]{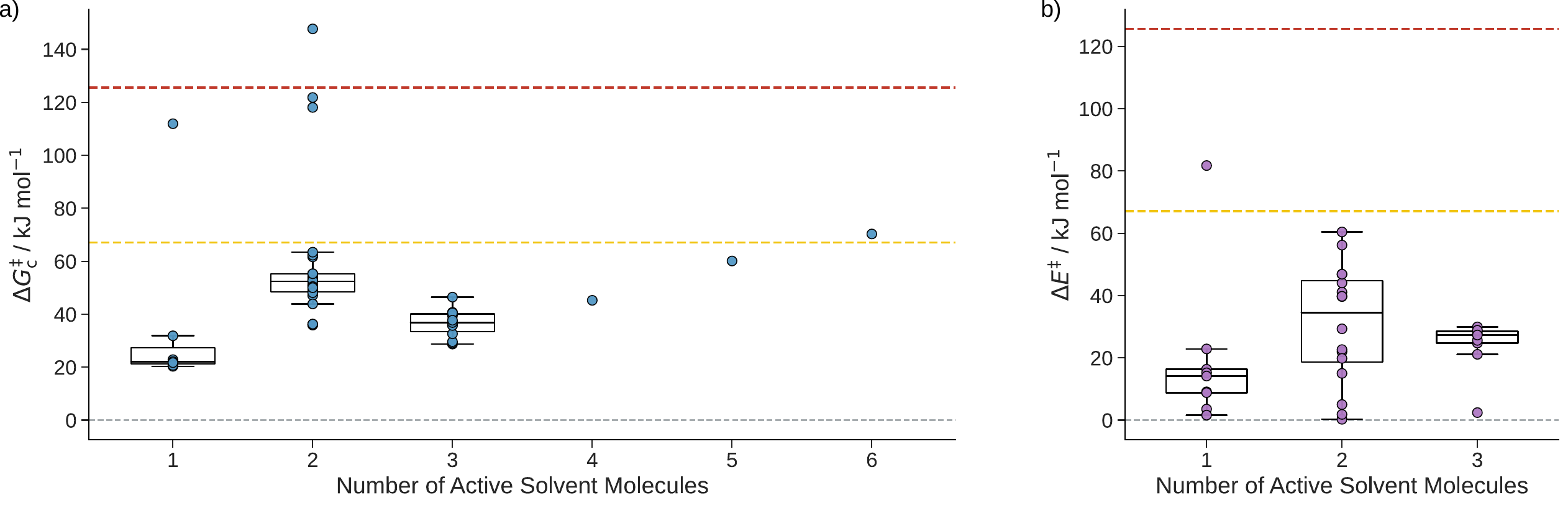}
  \caption{
    \textbf{a)} sQM free energies of activation $\Delta G^{\ddag}_\text{c}$ of the reaction of formaldehyde with water to form methanediol grouped by the number of active solvent molecules.
    \textbf{b)} mQM/MM energies of activation of the hydrolysis of formaldehyde grouped by the number of active solvent molecules.
    For groups with more than two data points, box plots are added where the whiskers of the boxes indicate the minimum and maximum values.
    The box borders depict the second and third quartile, and the line in a box denotes the median of the energies of the group.
    The dashed red line indicates the free energy of activation of \SI{125.6}{kJ~mol^{-1}} considering solely continuum solvation.
    The yellow dashed line indicates the experimental reference value of \SI{67.1}{kJ~mol^{-1}}.\cite{Winkelman2002}
  }
  \label{fig:resultsCH2O}
\end{figure}

With our scheme for the calculation of the free energies of activation $\Delta G^\ddag_\text{c}$, we now investigate the effect of actively involved solvent molecules.
The obtained results are grouped by the number of actively involved solvent molecules and shown in Fig.~\ref{fig:resultsCH2O}.
The grouping is based on the analysis of the eigenvector of the lowest vibrational eigenvalue of the Hessian of the TS with the algorithm described in section~\ref{sec:QMMMMEP}.

The four outliers around and above the activation energy derived from the continuum model have already been discussed above.
As shown in Fig.\ref{fig:resultsCH2O}a), mainly two, in some cases three, solvent molecules are actively involved in the TS of some configurations.
TS structures where two water molecules are involved form a 6-membered ring, where three involved water molecules form a 8-membered ring, which are reported as the key solvation motifs in the literature.\cite{Williams1983,Wolfe1995,Zhang2008,Inaba2015,Wang2023}
According to our results, the TS with three involved water molecules is preferred by $\Delta\Delta G^\ddag_\text{c}$=\SI{-17}{kJ~mol^{-1}}.
Configurations where four to six water molecules are actively involved turned out to be rare (three were found).
Other water molecules simply formed stabilizing hydrogen bonds with formaldehyde, but were not actively involved in the hydrolysis.

\begin{figure}[h]
	\centering
	\includegraphics[width=0.65\textwidth]{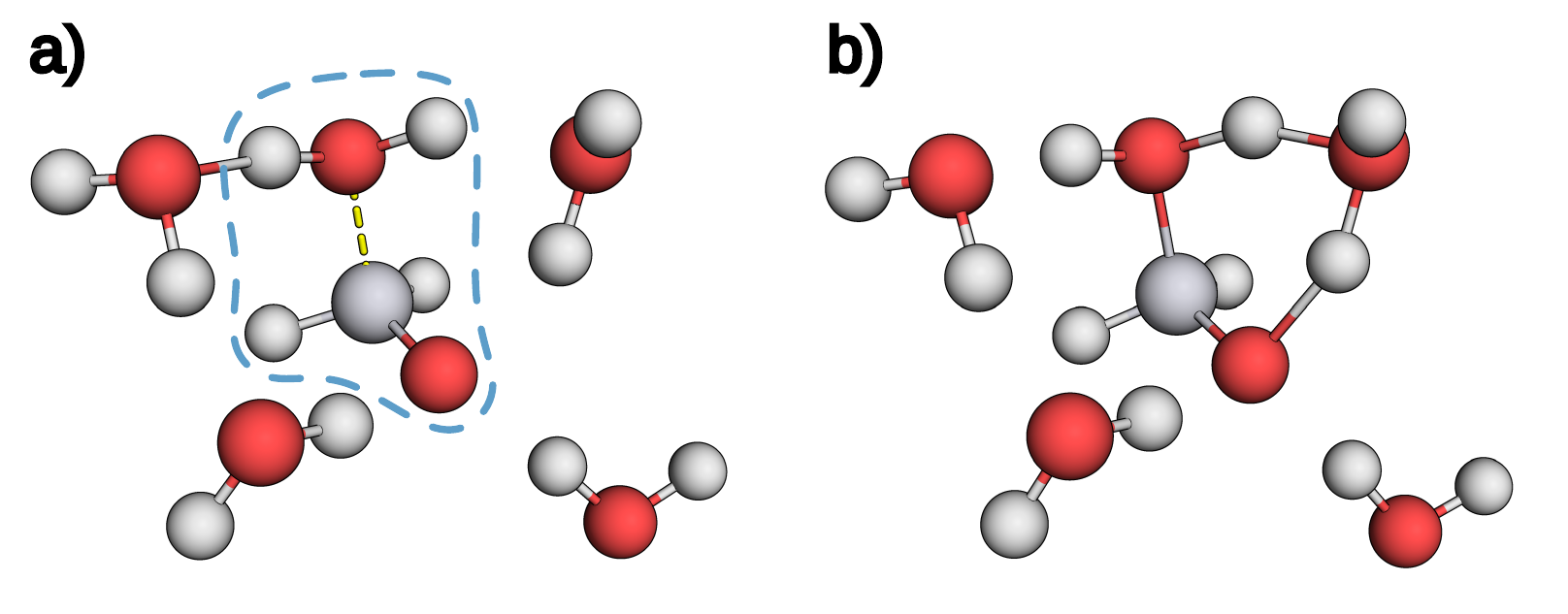}
	\caption{
		\textbf{a)} Local minimum structure of a sQM MEP with five water molecules.
		The dashed blue line encircles the motif considered to be bound due to stabilizing hydrogen bonds.
		The bond is indicated with a dashed yellow line.
		\textbf{b)} The TS structure of a sQM MEP with five water molecules.
                Two water molecules are actively involved and form a 6-membered ring with formaldehyde through the addition of a hydrogen atom to the oxygen atom in the carbonyl moiety.
	}
	\label{fig:CH2O_artificat}
\end{figure}

The low sQM free energies where only one solvent molecule seems to be involved appear to be artifacts of a strongly stabilized reactive complex with 5 to 7 solvent molecules (cf. supporting information Figure~S3) where one water molecule is considered to be already bound to formaldehyde, as depicted in Fig.~\ref{fig:CH2O_artificat}.
Two solvent molecules can be considered actively involved, as can be seen in the TS of such a configuration shown in Fig.~\ref{fig:CH2O_artificat}b).
Removing the strongly interacting water molecule on the left hand side of Fig.~\ref{fig:CH2O_artificat}b resolves such artifacts in the final decomposition to individual compounds.
Hence, the assembly of the minimum structures of the affected MEPs can be considered to consist of one formaldehyde molecule and 5 to 7 water molecules.

Considering the energies of activation $\Delta E^{\ddag}_\mathrm{el}$
obtained from the mQM/MM model, as shown in Fig.~\ref{fig:resultsCH2O}b), where no free energy corrections are included, these results are underestimated and their spread is too large for an accurate energy estimation.
This shows the requirement of our last step in the \textsc{Kingfisher} structure preparation model that reduces the degrees of freedom of our system and capture as much as possible of solvation effects with robust and reliable dielectric continuum models and more accurate electronic structure models.

\subsubsection{Chlorination of Phenol}
\label{sec:phenol}

\begin{figure}[!htb]
  \centering
  \includegraphics[width=0.55\textwidth]{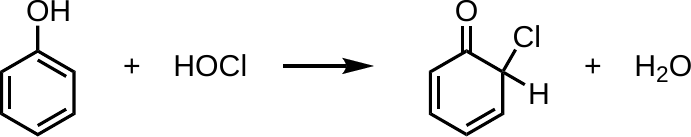}
  \caption{
    The formation of the deprotonated $\sigma$-complex during the chlorination of phenol with hypochlorous acid, represented by Lewis structures.
  }
  \label{fig:kfpPhenol}
\end{figure}

We analyzed the formation of the deprotonated $\sigma$-complex, simply referred to as $\sigma$-complex, in a reaction of phenol and hypochlorous acid in water.
The $\sigma$-complex is an intermediate in the mechanism of the electrophilic aromatic substitution of phenol to 2-chlorophenol.
The substitution reaction is of interest in water disinfection research as phenol is often employed as a surrogate model for natural organic matter.\cite{Bond2012,Prasse2020,Mazur2022}
In contrast to the previous example, water is now not a given as a reactant of this reaction.
Naturally, the reaction could not be found with a dielectric continuum solvation model only.
Clearly, water is not an innocent observer and explicit solvent molecules are required to obtain a MEP connecting phenol and its chlorinated, dearomatized and deprotonated $\sigma$-complex.

\begin{figure}[H]
  \centering
  \includegraphics[width=0.95\textwidth]{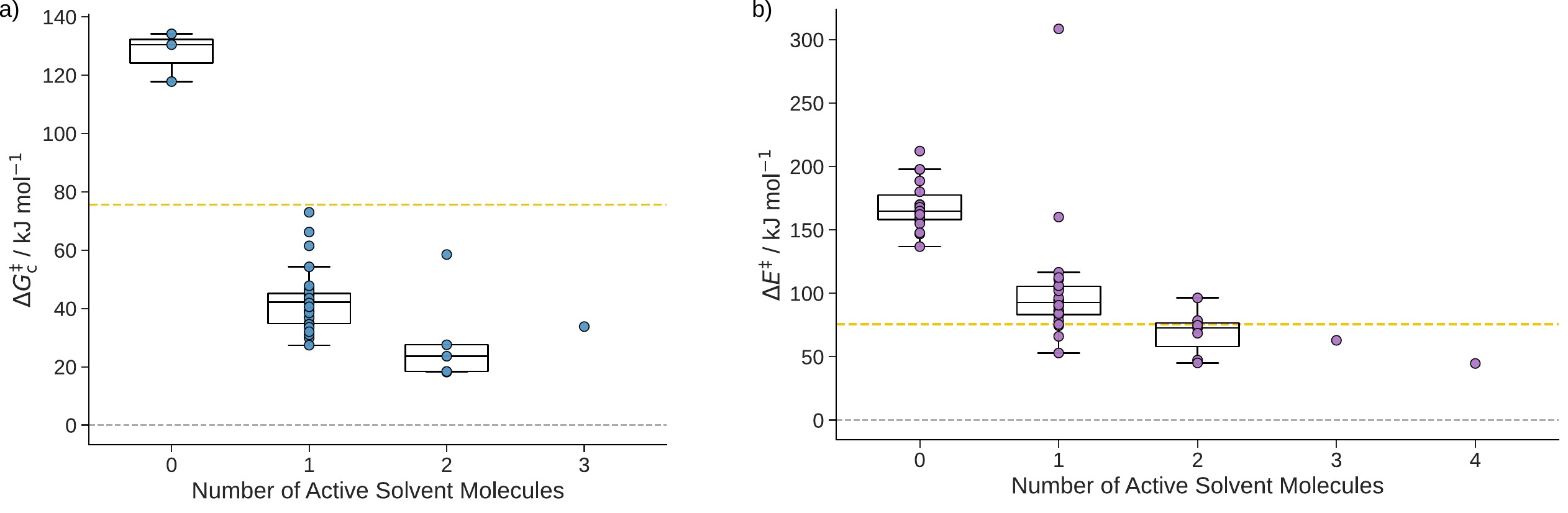}
  \caption{
    \textbf{a)} sQM free energies of activation $\Delta G^{\ddag}_\text{c}$ for the $\sigma$-complex formation from phenol in water sorted according to the number of active solvent molecules found in different configurations.
    \textbf{b)} mQM/MM energies of activation $\Delta E^{\ddag}_\text{el}$ of the $\sigma$-complex formation arranged in the same manner.
    For sets with more than two data points, box plots are added where the whiskers of the boxes indicate the minimum and maximum values, disregarding outliers.
    The box borders depict the second and third quartile, and the line in each box denotes the median of the energies of a group.
    The yellow dashed line indicates the experimental reference value of the chlorination of phenol \SI{75.6(19)}{kJ~mol^{-1}}.\cite{Gallard2002}
  }
  \label{fig:resultsPhenolHOCl}
\end{figure}

With roughly 200 reaction trials starting from different initial placements of solvent molecules, \textsc{Kingfisher} found 52 reactions within the mQM region and 36 within the sQM region (a comparison of energies of activation for this example is given in the supporting information).
The results presented in Fig.~\ref{fig:resultsPhenolHOCl} were analyzed and grouped with a minimum contribution threshold of $0.85\ c_\text{min}$ (cf. section~\ref{sec:QMMMMEP}).

In the obtained sQM MEPs, the number of total solvent molecules was in the range from two to five, while the number of active solvent molecules was zero to three.
Reactions where no solvent molecules are involved correspond to TS structures where the \ce{Cl-O} bond of hypochlorous acid is parallel to the plane of the aromatic ring.
In other TSs of this reaction,
the bond is perpendicular to this plane.
Hence, the relative orientation of the acid to the phenol determines whether the TS is favorable in energy or not, regardless of the solvent molecules in the model.

Most of the obtained MEPs
proceed with one solvent molecule involved.
The water molecule abstracts the hydrogen atom of the hydroxide group of phenol while adding one of its hydrogen atoms to the \ce{OH} of hypochlorous acid, forming an 8-membered ring.
The free energies of activation $\Delta G^{\ddag}_\text{c}$ for two solvent molecules involved are on average \SI{18}{kJ~mol^{-1}} lower in energy than with only one involved.
The reaction with three involved solvent molecules was sampled only once.

In the MEPs obtained for mQM TS structures, the number of solvents in the mQM region ranged from four to ten while subsequent analysis resulted in a range of active solvents from zero to only four.
With no active involvement of solvent molecules, the QM/MM energy of activation $\Delta E^{\ddag}_\text{el}$ is about \SI{75}{kJ~mol^{-1}} higher than the energy with an active involvement.
The trend depicted in Figure~\ref{fig:resultsPhenolHOCl}b) shows that more active solvent molecules result in lower energies of activation.
This indicates the significance of the active involvement of solvent molecules beyond simple stabilization by hydrogen bonds, somewhat regardless of the underlying electronic structure model.

Compared to the experimental free energy of activation,\cite{Gallard2002} the resulting free energies of activation $\Delta G^{\ddag}_\text{c}$ shown in Figure~\ref{fig:resultsPhenolHOCl}a) are again lower for reactions where solvents are actively involved, on average over all data points by \SI{36}{kJ~mol^{-1}}.
Note, however, that the experimental reference refers to chlorination of phenol in a neutral medium, whereas our data are just specific for the $\sigma$-complex formation.
The subsequent re-aromatization is assumed to be fast and the $\sigma$-complex formation to be the rate determining step.
Moreover, as for the previous example (see section~\ref{sec:ch2o}), the entropy loss is not fully captured by our approach.

\subsection{Hydration of \ce{CO2}}
\label{sec:co2results}

\begin{figure}[!htb]
  \centering
  \includegraphics[width=0.55\textwidth]{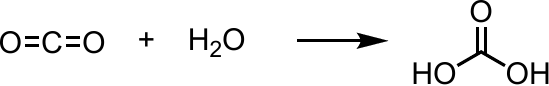}
  \caption{
    The hydration of carbon dioxide, represented by Lewis structures.
  }
  \label{fig:kfpCO2}
\end{figure}

The hydration of carbon dioxide (\ce{CO2}) in water and the resulting formation of carbonic acid has been investigated with molecular dynamics\cite{Leung2007,Kumar2009,Stirling2010,Gallet2012,Martirez2023}, but also with time-independent approaches\cite{Nguyen1997,Nguyen2008,Wang2013} approaches and by experiment.\cite{Adamczyk2009,Wang2010,England2011}
The reaction attracts attention due to the acidification of oceans through increasing atmospheric \ce{CO2} concentration.\cite{Honisch2012}
Despite being polarized, \ce{CO2} has no dipole moment and interacts poorly with water.
Hence, its hydration
offers an ideal challenge for \textsc{Kingfisher}.
We applied \textsc{Kingfisher} in a setting of pure water as well as in a one-to-one mixture of water and methanol to illustrate its general applicability.

\begin{figure}[H]
  \centering
  \includegraphics[width=0.95\textwidth]{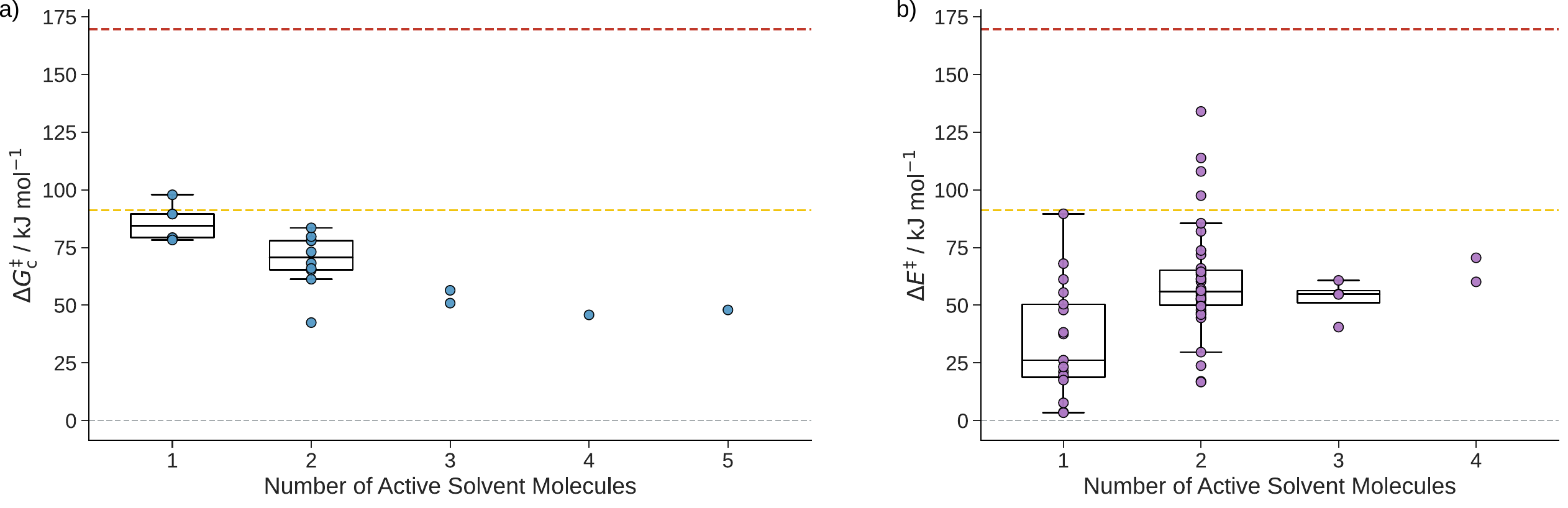}
  \caption{
    \textbf{a)} sQM free energies of activation $\Delta G^{\ddag}_\text{c}$ of the hydration of \ce{CO2} grouped by the number of active solvent molecules.
    \textbf{b)} The mQM/MM energies of activation $\Delta E^{\ddag}_\text{el}$ of the hydration of \ce{CO2} grouped according to the number of active solvent molecules.
    For groups with more than two data points, box plots are added where the whiskers of the boxes indicate the minimum and maximum values, disregarding outliers.
    The box borders depict the second and third quartile, and the line in each box denotes the median of the energies of a group.
    The dashed red line indicates the free energy of activation of \SI{169.6}{kJ~mol^{-1}} only considering continuum solvation.
    The yellow dashed line indicates the experimental reference value of \SI{90.2}{kJ~mol^{-1}}.\cite{Wang2010}
  }
  \label{fig:resultsCO2}
\end{figure}

With about 200 reaction trials starting from different initial placements of water molecules, \textsc{Kingfisher} automatically found 58 reactions within the mQM and 20 within the sQM region and their respective electronic structure methods.
The comparison of the energies of activation can be found in the supporting information.
The results presented in Fig.~\ref{fig:resultsCO2} were analyzed and grouped as described in section~\ref{sec:ch2o}.

For the sQM reactions (Fig.~\ref{fig:resultsCO2}a), systems with two, three, four, and one with six solvent molecules were found, one to five of them were actively involved.
For the four cases where one solvent molecule appeared to be involved, we found, upon visual inspection and variation of $c_\text{min}$ for this analysis (by using a percentage of it as threshold for the algorithm presented in section~\ref{sec:QMMMMEP}), that two solvent molecules are involved in theses cases.
By adapting $c_\text{min}$ for the data analysis only as described in section~\ref{sec:phenol}, they could be re-assigned to another group, but in order to illustrate the limitations of the \textsc{Kingfisher} algorithm we did not chose to do so.
The majority of reactions involved two solvent molecules.

Analyzing the results of the mQM MEPs shown in Fig.~\ref{fig:resultsCO2}b), we find a similar trend as observed for formaldehyde.
The mQM/MM energy of activation slightly increases with the number of involved solvent molecules and has a large spread.
Over all data points for free energies of activation obtained with the sQM model, the experimental free energy of activation is underestimated by \SI{20}{kJ~mol^{-1}}.
The entropy and enthalpy of activation have been determined experimentally.\cite{Wang2010}
In contrast to our previous examples, \textsc{Kingfisher} now overestimates the loss in entropy, where the experimental entropy of activation is \SI{-42}{J~mol^{-1}~K^{-1}} and our result, with $\Delta S^\ddag + \Delta S^\ddag_\text{c}$, is \SI{-107(24)}{J~mol^{-1}~K^{-1}}.
However, with an enthalpy of activation of \SI{39(14)}{kJ~mol^{-1}} \textsc{Kingfisher} underestimates the experimental value of \SI{79}{kJ~mol^{-1}}.
Hence, the good agreement with the experimental results originates from fortunate error cancellation.
It is hypothesized that the error in the enthalpy of activation, or the electronic energy of activation (compare Eq.~\eqref{eq:Hact}), might be due to shortcomings of DFT in properly describing the change in polarity of \ce{CO2} when approaching the TS.\cite{Martirez2023}

\subsubsection{Hydration of \ce{CO2} in a Water/Methanol Mixture}

\begin{figure}[!htb]
  \centering
  \includegraphics[width=0.95\textwidth]{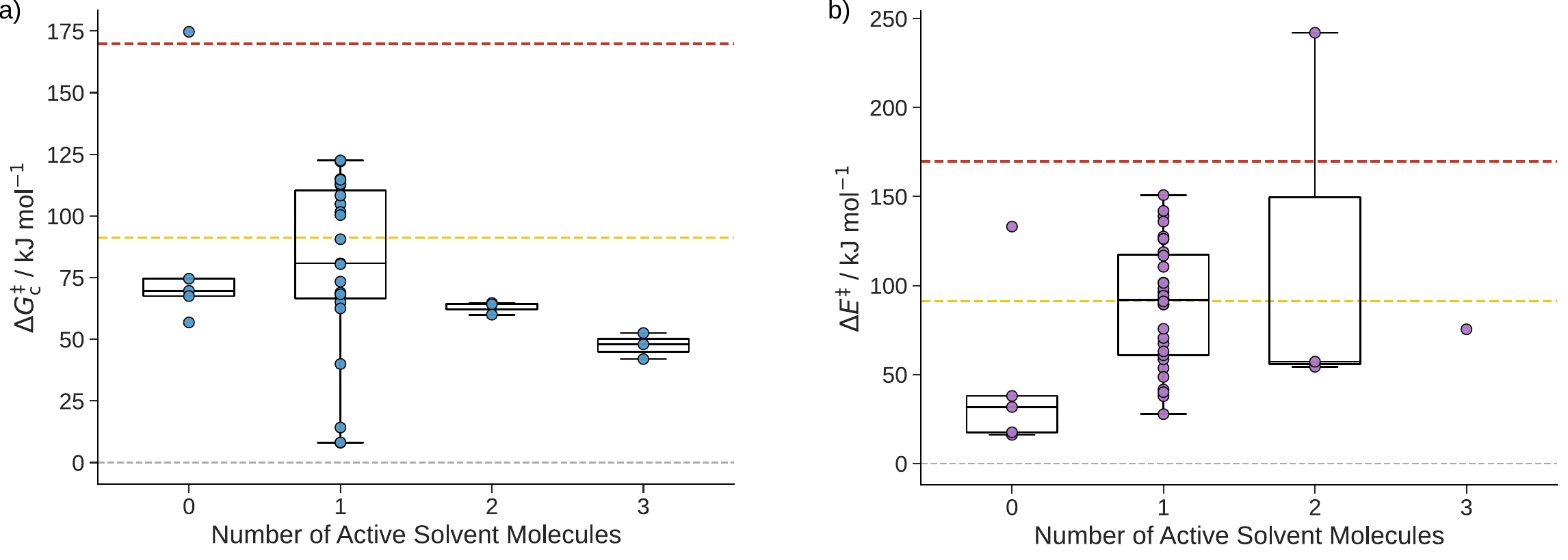}
  \caption{
    \textbf{a)} sQM free energies of activation $\Delta G^{\ddag}_\text{c}$ of the hydration of \ce{CO2} in a water/methanol mixture grouped by the number of active solvent molecules.
    \textbf{b)} mQM/MM energies of activation $\Delta E^{\ddag}_\text{el}$ of the hydration of \ce{CO2} in a water/methanol mixture sorted according to the number of active solvent molecules.
    For groups with more than two data points, box plots are added where the whiskers of the boxes indicate the minimum and maximum values, disregarding outliers.
    The box borders depict the second and third quartile, and the line in each box denotes the median of the energies of a group.
    The dashed red line indicates the free energy of activation of \SI{169.8}{kJ~mol^{-1}} only considering continuum solvation.
    The yellow dashed line indicates the experimental reference value in pure water of \SI{90.2}{kJ~mol^{-1}}.\cite{Wang2010}
  }
  \label{fig:resultsCO2inMix}
\end{figure}

To present the capabilities of \textsc{Kingfisher}, we investigated the hydration
of \ce{CO2} in a $1:1$ mixture of water and methanol,
with one water molecule being considered to be a reactant.
To do so, the average volume of both solvents was calculated for Eq.\eqref{eq:radiiRatio} and the relative permittivity approximated to be $56.52$, the average permittivity of the pure solvents.
In view of these assumptions, we do not distinguish between the type of solvents in the following analysis.
\textsc{Kingfisher} identified 41 reactions within the mQM and 34 within the sQM region.
The free energies of activation grouped by the number of involved solvent molecules are shown in Fig.~\ref{fig:resultsCO2inMix}.

Since one water molecule was considered a reactant, the examples with zero active solvent molecules means that no other solvent molecules were involved or assisted actively in the reaction.
The outlier above the dashed red line in Fig.~\ref{fig:resultsCO2inMix}a), indicating the free energy of activation in dielectric continuum solvation, corresponds to the
reference TS in the continuum.
The barrier is slightly higher due to the correction term for the cavity entropy.

The remaining examples with no active involvement of solvent molecules are assisted by methanol via a proton shuttle and one example shows an assisted proton transfer from the reacting water molecule.
For the results with one solvent molecule actively involved, the three cases with lowest free energies of activation form strongly stabilized minimum structures with the reacting water molecule, similar to the phenomena described in section~\ref{sec:ch2o}.
Overall, the free energy of activation $\Delta G^{\ddag}_\text{c}$ decreases with the number of solvent molecules involved, the average energy of activation is slightly higher compared to the hydration in pure water.

\begin{figure}[H]
    \centering
    \label{fig:co2_TScomparison}
    \includegraphics[width=0.65\textwidth]{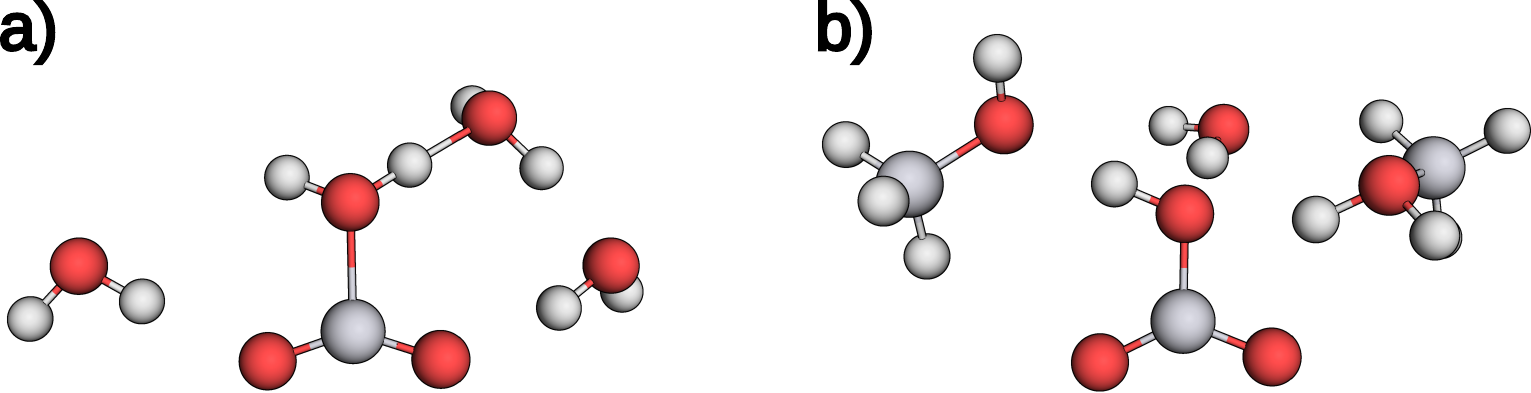}
    \caption{
        \textbf{a)} One sQM TS structure of the hydrolysis of \ce{CO2} with four water molecules.
        Three water molecules are actively involved and form a 8-membered ring with \ce{CO2}.
        \textbf{b)} One sQM TS structure of the hydrolysis of \ce{CO2} with two water and two methanol molecules.
        Two solvent molecules are actively involved in the formation of carbonic acid,
        one methanol molecule helping in the transfer of the added water molecule.
    }
\end{figure}

Results with energies higher than the experimental reference barrier (measured in pure water) highlight a key difference from the calculated energies discussed for the hydration of \ce{CO2} in pure water.
The TSs are higher in energy, if the reacting water molecule transfers one of its hydrogen atoms directly to an oxygen atom of \ce{CO2}, assisted by a solvent molecule.
Such an arrangement is higher in energy than the formation of a 6-membered ring with a proton shuffle.
One example for each case is shown in Fig.~\ref{fig:co2_TScomparison}.
These TSs appear to be found more frequently in a solvent mixture with lower relative permittivity than water.
Considering the mQM results in Fig.~\ref{fig:resultsCO2inMix}b), the spread of the mQM/MM energies of activation is large, between MEPs with one to six solvent molecules are found.

\section{Conclusion}
In this work, we presented the structure preparation model \textsc{Kingfisher} which allows for a fully automated identification of transition state structures with active solvent molecules for any reaction and any solvent or solvent mixture.
\textsc{Kingfisher} is capable of producing transition state motifs with minimal solvent-molecule contribution.
A key ingredient in the procedure, that allows \textsc{Kingfisher} to boil down a large microsolvated QM/MM hybrid model to such a minimal microsolvated-continuum model with active solvent molecules, is the analysis of the decaying normal mode at the transition state with respect to solvent-molecule contributions. 

To supplement the structures derived with free-energy data, we extended the canonical quantum chemical free energy model for the gas phase by a cavity entropy term to obtain an easy-to-evaluate solvation free energy approach.
We investigated the algorithms for three different reactions, for which \textsc{Kingfisher} isolated solvent-molecule contributions to transition states. The free-energy results obtained agree fairly well with experimental reference data.

\section*{Data Availability}
\label{sec:data}
All reaction networks presented in this paper are available on Zenodo.\cite{kingfisher2024}
They are stored in our MongoDB framework with a description on how to extract them.
Additionally, we have added all resources and scripts to reproduce all results with our \textsc{SCINE} software framework, including the Apptainer container of the \textsc{SCINE Puffin} that executed all calculations, as well as the scripts required for data analysis and visualization.

\section*{Acknowledgments}
\label{sec:acknowledgments}
The authors gratefully acknowledge financial support through ETH grant ETH-44~20-1.


\begin{mcitethebibliography}{121}
\providecommand*\natexlab[1]{#1}
\providecommand*\mciteSetBstSublistMode[1]{}
\providecommand*\mciteSetBstMaxWidthForm[2]{}
\providecommand*\mciteBstWouldAddEndPuncttrue
  {\def\EndOfBibitem{\unskip.}}
\providecommand*\mciteBstWouldAddEndPunctfalse
  {\let\EndOfBibitem\relax}
\providecommand*\mciteSetBstMidEndSepPunct[3]{}
\providecommand*\mciteSetBstSublistLabelBeginEnd[3]{}
\providecommand*\EndOfBibitem{}
\mciteSetBstSublistMode{f}
\mciteSetBstMaxWidthForm{subitem}{(\alph{mcitesubitemcount})}
\mciteSetBstSublistLabelBeginEnd
  {\mcitemaxwidthsubitemform\space}
  {\relax}
  {\relax}

\bibitem[Barone and Cossi(1998)Barone, and Cossi]{Barone1998}
Barone,~V.; Cossi,~M. {Quantum Calculation of Molecular Energies and Energy
  Gradients in Solution by a Conductor Solvent Model}. \emph{J. Phys. Chem. A}
  \textbf{1998}, \emph{102}, 1995--2001\relax
\mciteBstWouldAddEndPuncttrue
\mciteSetBstMidEndSepPunct{\mcitedefaultmidpunct}
{\mcitedefaultendpunct}{\mcitedefaultseppunct}\relax
\EndOfBibitem
\bibitem[Tomasi \latin{et~al.}(2005)Tomasi, Mennucci, and Cammi]{Tomasi2005}
Tomasi,~J.; Mennucci,~B.; Cammi,~R. {Quantum Mechanical Continuum Solvation
  Models}. \emph{Chem. Rev.} \textbf{2005}, \emph{105}, 2999--3094\relax
\mciteBstWouldAddEndPuncttrue
\mciteSetBstMidEndSepPunct{\mcitedefaultmidpunct}
{\mcitedefaultendpunct}{\mcitedefaultseppunct}\relax
\EndOfBibitem
\bibitem[Mennucci(2012)]{Mennucci2012}
Mennucci,~B. {Polarizable Continuum Model}. \emph{WIREs Comput Mol Sci.}
  \textbf{2012}, \emph{2}, 386--404\relax
\mciteBstWouldAddEndPuncttrue
\mciteSetBstMidEndSepPunct{\mcitedefaultmidpunct}
{\mcitedefaultendpunct}{\mcitedefaultseppunct}\relax
\EndOfBibitem
\bibitem[Herbert(2021)]{Herbert2021}
Herbert,~J.~M. {Dielectric Continuum Methods for Quantum Chemistry}.
  \emph{WIREs Comput Mol Sci.} \textbf{2021}, \emph{11}, e1519\relax
\mciteBstWouldAddEndPuncttrue
\mciteSetBstMidEndSepPunct{\mcitedefaultmidpunct}
{\mcitedefaultendpunct}{\mcitedefaultseppunct}\relax
\EndOfBibitem
\bibitem[Klamt(2018)]{Klamt2018}
Klamt,~A. {The COSMO and COSMO-RS Solvation Models}. \emph{WIREs Comput. Mol.
  Sci.} \textbf{2018}, \emph{8}, e1338\relax
\mciteBstWouldAddEndPuncttrue
\mciteSetBstMidEndSepPunct{\mcitedefaultmidpunct}
{\mcitedefaultendpunct}{\mcitedefaultseppunct}\relax
\EndOfBibitem
\bibitem[Norjmaa \latin{et~al.}(2022)Norjmaa, Ujaque, and
  Lled{\'o}s]{Norjmaa2022}
Norjmaa,~G.; Ujaque,~G.; Lled{\'o}s,~A. {Beyond Continuum Solvent Models in
  Computational Homogeneous Catalysis}. \emph{Top. Catal.} \textbf{2022},
  \emph{65}, 118--140\relax
\mciteBstWouldAddEndPuncttrue
\mciteSetBstMidEndSepPunct{\mcitedefaultmidpunct}
{\mcitedefaultendpunct}{\mcitedefaultseppunct}\relax
\EndOfBibitem
\bibitem[Pliego and Riveros(2020)Pliego, and Riveros]{Pliego2020}
Pliego,~J.~R.; Riveros,~J.~M. {Hybrid Discrete-Continuum Solvation Methods}.
  \emph{WIREs Comput Mol Sci.} \textbf{2020}, \emph{10}, e1440\relax
\mciteBstWouldAddEndPuncttrue
\mciteSetBstMidEndSepPunct{\mcitedefaultmidpunct}
{\mcitedefaultendpunct}{\mcitedefaultseppunct}\relax
\EndOfBibitem
\bibitem[Sunoj and Anand(2012)Sunoj, and Anand]{Sunoj2012}
Sunoj,~R.~B.; Anand,~M. {Microsolvated Transition State Models for Improved
  Insight into Chemical Properties and Reaction Mechanisms}. \emph{Phys. Chem.
  Chem. Phys.} \textbf{2012}, \emph{14}, 12715--12736\relax
\mciteBstWouldAddEndPuncttrue
\mciteSetBstMidEndSepPunct{\mcitedefaultmidpunct}
{\mcitedefaultendpunct}{\mcitedefaultseppunct}\relax
\EndOfBibitem
\bibitem[Basdogan \latin{et~al.}(2020)Basdogan, Maldonado, and
  Keith]{Basdogan2020a}
Basdogan,~Y.; Maldonado,~A.~M.; Keith,~J.~A. {Advances and Challenges in
  Modeling Solvated Reaction Mechanisms for Renewable Fuels and Chemicals}.
  \emph{WIREs Comput Mol Sci.} \textbf{2020}, \emph{10}, e1446\relax
\mciteBstWouldAddEndPuncttrue
\mciteSetBstMidEndSepPunct{\mcitedefaultmidpunct}
{\mcitedefaultendpunct}{\mcitedefaultseppunct}\relax
\EndOfBibitem
\bibitem[Das \latin{et~al.}(2022)Das, Gogoi, and Sunoj]{Das2022}
Das,~M.; Gogoi,~A.~R.; Sunoj,~R.~B. {Molecular Insights on Solvent Effects in
  Organic Reactions as Obtained through Computational Chemistry Tools}.
  \emph{J. Org. Chem.} \textbf{2022}, \emph{87}, 1630--1640\relax
\mciteBstWouldAddEndPuncttrue
\mciteSetBstMidEndSepPunct{\mcitedefaultmidpunct}
{\mcitedefaultendpunct}{\mcitedefaultseppunct}\relax
\EndOfBibitem
\bibitem[Warshel and Levitt(1976)Warshel, and Levitt]{Warshel1976a}
Warshel,~A.; Levitt,~M. {Theoretical Studies of Enzymic Reactions: Dielectric,
  Electrostatic and Steric Stabilization of the Carbonium Ion in the Reaction
  of Lysozyme}. \emph{J. Mol. Biol.} \textbf{1976}, \emph{103}, 227--249\relax
\mciteBstWouldAddEndPuncttrue
\mciteSetBstMidEndSepPunct{\mcitedefaultmidpunct}
{\mcitedefaultendpunct}{\mcitedefaultseppunct}\relax
\EndOfBibitem
\bibitem[Field \latin{et~al.}(1990)Field, Bash, and Karplus]{Field1990}
Field,~M.~J.; Bash,~P.~A.; Karplus,~M. {A Combined Quantum Mechanical and
  Molecular Mechanical Potential for Molecular Dynamics Simulations}. \emph{J.
  Comput. Chem.} \textbf{1990}, \emph{11}, 700--733\relax
\mciteBstWouldAddEndPuncttrue
\mciteSetBstMidEndSepPunct{\mcitedefaultmidpunct}
{\mcitedefaultendpunct}{\mcitedefaultseppunct}\relax
\EndOfBibitem
\bibitem[Lipparini and Mennucci(2021)Lipparini, and Mennucci]{Lipparini2021}
Lipparini,~F.; Mennucci,~B. {Hybrid QM/Classical Models: Methodological
  Advances and New Applications}. \emph{Chem. Phys. Rev.} \textbf{2021},
  \emph{2}, 041303\relax
\mciteBstWouldAddEndPuncttrue
\mciteSetBstMidEndSepPunct{\mcitedefaultmidpunct}
{\mcitedefaultendpunct}{\mcitedefaultseppunct}\relax
\EndOfBibitem
\bibitem[Csizi and Reiher(2023)Csizi, and Reiher]{Csizi2023a}
Csizi,~K.-S.; Reiher,~M. {Universal QM/MM Approaches for General Nanoscale
  Applications}. \emph{WIREs Comput Mol Sci.} \textbf{2023}, \emph{13},
  e1656\relax
\mciteBstWouldAddEndPuncttrue
\mciteSetBstMidEndSepPunct{\mcitedefaultmidpunct}
{\mcitedefaultendpunct}{\mcitedefaultseppunct}\relax
\EndOfBibitem
\bibitem[Demapan \latin{et~al.}(2022)Demapan, Kussmann, Ochsenfeld, and
  Cui]{Demapan2022}
Demapan,~D.; Kussmann,~J.; Ochsenfeld,~C.; Cui,~Q. {Factors That Determine the
  Variation of Equilibrium and Kinetic Properties of QM/MM Enzyme Simulations:
  QM Region, Conformation, and Boundary Condition}. \emph{J. Chem. Theory
  Comput.} \textbf{2022}, \emph{18}, 2530--2542\relax
\mciteBstWouldAddEndPuncttrue
\mciteSetBstMidEndSepPunct{\mcitedefaultmidpunct}
{\mcitedefaultendpunct}{\mcitedefaultseppunct}\relax
\EndOfBibitem
\bibitem[Ho \latin{et~al.}(2024)Ho, Yu, Shao, Taylor, and Chen]{Ho2024}
Ho,~J.; Yu,~H.; Shao,~Y.; Taylor,~M.; Chen,~J. {How Accurate Are QM/MM Models?}
  \emph{J. Phys. Chem. A} \textbf{2024}, \relax
\mciteBstWouldAddEndPunctfalse
\mciteSetBstMidEndSepPunct{\mcitedefaultmidpunct}
{}{\mcitedefaultseppunct}\relax
\EndOfBibitem
\bibitem[Chandrasekhar \latin{et~al.}(2002)Chandrasekhar, Shariffskul, and
  Jorgensen]{Chandrasekhar2002}
Chandrasekhar,~J.; Shariffskul,~S.; Jorgensen,~W.~L. {QM/MM Simulations for
  Diels-Alder Reactions in Water:\, Contribution of Enhanced Hydrogen Bonding
  at the Transition State to the Solvent Effect}. \emph{J. Phys. Chem. B}
  \textbf{2002}, \emph{106}, 8078--8085\relax
\mciteBstWouldAddEndPuncttrue
\mciteSetBstMidEndSepPunct{\mcitedefaultmidpunct}
{\mcitedefaultendpunct}{\mcitedefaultseppunct}\relax
\EndOfBibitem
\bibitem[Yang \latin{et~al.}(2015)Yang, Doubleday, and Houk]{Yang2015}
Yang,~Z.; Doubleday,~C.; Houk,~K.~N. {QM/MM Protocol for Direct Molecular
  Dynamics of Chemical Reactions in Solution: The Water-Accelerated
  Diels--Alder Reaction}. \emph{J. Chem. Theory Comput.} \textbf{2015},
  \emph{11}, 5606--5612\relax
\mciteBstWouldAddEndPuncttrue
\mciteSetBstMidEndSepPunct{\mcitedefaultmidpunct}
{\mcitedefaultendpunct}{\mcitedefaultseppunct}\relax
\EndOfBibitem
\bibitem[Boereboom \latin{et~al.}(2018)Boereboom, {Fleurat-Lessard}, and
  Bulo]{Boereboom2018}
Boereboom,~J.~M.; {Fleurat-Lessard},~P.; Bulo,~R.~E. {Explicit Solvation
  Matters: Performance of QM/MM Solvation Models in Nucleophilic Addition}.
  \emph{J. Chem. Theory Comput.} \textbf{2018}, \emph{14}, 1841--1852\relax
\mciteBstWouldAddEndPuncttrue
\mciteSetBstMidEndSepPunct{\mcitedefaultmidpunct}
{\mcitedefaultendpunct}{\mcitedefaultseppunct}\relax
\EndOfBibitem
\bibitem[Clabaut \latin{et~al.}(2020)Clabaut, Schweitzer, G{\"o}tz, Michel, and
  Steinmann]{Clabaut2020}
Clabaut,~P.; Schweitzer,~B.; G{\"o}tz,~A.~W.; Michel,~C.; Steinmann,~S.~N.
  {Solvation Free Energies and Adsorption Energies at the Metal/Water Interface
  from Hybrid Quantum-Mechanical/Molecular Mechanics Simulations}. \emph{J.
  Chem. Theory Comput.} \textbf{2020}, \emph{16}, 6539--6549\relax
\mciteBstWouldAddEndPuncttrue
\mciteSetBstMidEndSepPunct{\mcitedefaultmidpunct}
{\mcitedefaultendpunct}{\mcitedefaultseppunct}\relax
\EndOfBibitem
\bibitem[Feh{\'e}r and Stirling(2019)Feh{\'e}r, and Stirling]{Feher2019}
Feh{\'e}r,~P.~P.; Stirling,~A. {Assessment of Reactivities with Explicit and
  Implicit Solvent Models: QM/MM and Gas-Phase Evaluation of Three Different
  Ag-catalysed Furan Ring Formation Routes}. \emph{New J. Chem.} \textbf{2019},
  \emph{43}, 15706--15713\relax
\mciteBstWouldAddEndPuncttrue
\mciteSetBstMidEndSepPunct{\mcitedefaultmidpunct}
{\mcitedefaultendpunct}{\mcitedefaultseppunct}\relax
\EndOfBibitem
\bibitem[Meng \latin{et~al.}(2023)Meng, Zhang, Ramirez, and Ayers]{Meng2023}
Meng,~F.; Zhang,~H.; Ramirez,~J. S.~C.; Ayers,~P.~W. {Something for Nothing:
  Improved Solvation Free Energy Prediction with $\Delta$-Learning}.
  \emph{Theor. Chem. Acc.} \textbf{2023}, \emph{142}, 106\relax
\mciteBstWouldAddEndPuncttrue
\mciteSetBstMidEndSepPunct{\mcitedefaultmidpunct}
{\mcitedefaultendpunct}{\mcitedefaultseppunct}\relax
\EndOfBibitem
\bibitem[Fowles and Palmer(2023)Fowles, and Palmer]{Fowles2023}
Fowles,~D.; Palmer,~D.~S. {Solvation Entropy, Enthalpy and Free Energy
  Prediction Using a Multi-Task Deep Learning Functional in 1D-RISM}.
  \emph{Phys. Chem. Chem. Phys.} \textbf{2023}, \emph{25}, 6944--6954\relax
\mciteBstWouldAddEndPuncttrue
\mciteSetBstMidEndSepPunct{\mcitedefaultmidpunct}
{\mcitedefaultendpunct}{\mcitedefaultseppunct}\relax
\EndOfBibitem
\bibitem[Zhang \latin{et~al.}(2023)Zhang, Juraskova, and Duarte]{Zhang2023}
Zhang,~H.; Juraskova,~V.; Duarte,~F. {Modeling Chemical Processes in Explicit
  Solvents with Machine Learning Potentials}. \emph{ChemRxiv} \textbf{2023},
  DOI: 10.26434/chemrxiv-2023-ktscq\relax
\mciteBstWouldAddEndPuncttrue
\mciteSetBstMidEndSepPunct{\mcitedefaultmidpunct}
{\mcitedefaultendpunct}{\mcitedefaultseppunct}\relax
\EndOfBibitem
\bibitem[C{\'e}lerse \latin{et~al.}(2024)C{\'e}lerse, Juraskova, Das, Wodrich,
  and Corminboeuf]{Celerse2024}
C{\'e}lerse,~F.; Juraskova,~V.; Das,~S.; Wodrich,~M.~D.; Corminboeuf,~C.
  {Capturing Dichotomic Solvent Behavior in Solute--Solvent Reactions with
  Neural Network Potentials}. \emph{J. Chem. Theory Comput.} \textbf{2024},
  \emph{20}, 10350--10361\relax
\mciteBstWouldAddEndPuncttrue
\mciteSetBstMidEndSepPunct{\mcitedefaultmidpunct}
{\mcitedefaultendpunct}{\mcitedefaultseppunct}\relax
\EndOfBibitem
\bibitem[Varghese and Mushrif(2019)Varghese, and Mushrif]{Varghese2019}
Varghese,~J.~J.; Mushrif,~S.~H. {Origins of Complex Solvent Effects on Chemical
  Reactivity and Computational Tools to Investigate Them: A Review}.
  \emph{React. Chem. Eng.} \textbf{2019}, \emph{4}, 165--206\relax
\mciteBstWouldAddEndPuncttrue
\mciteSetBstMidEndSepPunct{\mcitedefaultmidpunct}
{\mcitedefaultendpunct}{\mcitedefaultseppunct}\relax
\EndOfBibitem
\bibitem[Sameera \latin{et~al.}(2016)Sameera, Maeda, and Morokuma]{Sameera2016}
Sameera,~W. M.~C.; Maeda,~S.; Morokuma,~K. {Computational Catalysis Using the
  Artificial Force Induced Reaction Method}. \emph{Acc. Chem. Res.}
  \textbf{2016}, \emph{49}, 763--773\relax
\mciteBstWouldAddEndPuncttrue
\mciteSetBstMidEndSepPunct{\mcitedefaultmidpunct}
{\mcitedefaultendpunct}{\mcitedefaultseppunct}\relax
\EndOfBibitem
\bibitem[Dewyer and Zimmerman(2017)Dewyer, and Zimmerman]{Dewyer2017}
Dewyer,~A.~L.; Zimmerman,~P.~M. {Finding Reaction Mechanisms, Intuitive or
  Otherwise}. \emph{Org. Biomol. Chem.} \textbf{2017}, \emph{15},
  501--504\relax
\mciteBstWouldAddEndPuncttrue
\mciteSetBstMidEndSepPunct{\mcitedefaultmidpunct}
{\mcitedefaultendpunct}{\mcitedefaultseppunct}\relax
\EndOfBibitem
\bibitem[Simm \latin{et~al.}(2019)Simm, Vaucher, and Reiher]{Simm2019}
Simm,~G.~N.; Vaucher,~A.~C.; Reiher,~M. {Exploration of Reaction Pathways and
  Chemical Transformation Networks}. \emph{J. Phys. Chem. A} \textbf{2019},
  \emph{123}, 385--399\relax
\mciteBstWouldAddEndPuncttrue
\mciteSetBstMidEndSepPunct{\mcitedefaultmidpunct}
{\mcitedefaultendpunct}{\mcitedefaultseppunct}\relax
\EndOfBibitem
\bibitem[Unsleber and Reiher(2020)Unsleber, and Reiher]{Unsleber2020}
Unsleber,~J.~P.; Reiher,~M. {The Exploration of Chemical Reaction Networks}.
  \emph{Annu. Rev. Phys. Chem.} \textbf{2020}, \emph{71}, 121--142\relax
\mciteBstWouldAddEndPuncttrue
\mciteSetBstMidEndSepPunct{\mcitedefaultmidpunct}
{\mcitedefaultendpunct}{\mcitedefaultseppunct}\relax
\EndOfBibitem
\bibitem[Steiner and Reiher(2022)Steiner, and Reiher]{Steiner2022}
Steiner,~M.; Reiher,~M. {Autonomous Reaction Network Exploration in Homogeneous
  and Heterogeneous Catalysis}. \emph{Top. Catal.} \textbf{2022}, \emph{65},
  6--39\relax
\mciteBstWouldAddEndPuncttrue
\mciteSetBstMidEndSepPunct{\mcitedefaultmidpunct}
{\mcitedefaultendpunct}{\mcitedefaultseppunct}\relax
\EndOfBibitem
\bibitem[Baiardi \latin{et~al.}(2022)Baiardi, Grimmel, Steiner, T{\"u}rtscher,
  Unsleber, Weymuth, and Reiher]{Baiardi2022}
Baiardi,~A.; Grimmel,~S.~A.; Steiner,~M.; T{\"u}rtscher,~P.~L.;
  Unsleber,~J.~P.; Weymuth,~T.; Reiher,~M. {Expansive Quantum Mechanical
  Exploration of Chemical Reaction Paths}. \emph{Acc. Chem. Res.}
  \textbf{2022}, \emph{55}, 35--43\relax
\mciteBstWouldAddEndPuncttrue
\mciteSetBstMidEndSepPunct{\mcitedefaultmidpunct}
{\mcitedefaultendpunct}{\mcitedefaultseppunct}\relax
\EndOfBibitem
\bibitem[Ismail \latin{et~al.}(2022)Ismail, Majerus, and Habershon]{Ismail2022}
Ismail,~I.; Majerus,~R.~C.; Habershon,~S. {Graph-Driven Reaction Discovery:
  Progress, Challenges, and Future Opportunities}. \emph{J. Phys. Chem. A}
  \textbf{2022}, \emph{126}, 7051--7069\relax
\mciteBstWouldAddEndPuncttrue
\mciteSetBstMidEndSepPunct{\mcitedefaultmidpunct}
{\mcitedefaultendpunct}{\mcitedefaultseppunct}\relax
\EndOfBibitem
\bibitem[Wen \latin{et~al.}(2023)Wen, Spotte-Smith, Blau, McDermott,
  Krishnapriyan, and Persson]{Wen2023}
Wen,~M.; Spotte-Smith,~E. W.~C.; Blau,~S.~M.; McDermott,~M.~J.;
  Krishnapriyan,~A.~S.; Persson,~K.~A. {Chemical reaction networks and
  opportunities for machine learning}. \emph{Nat. Comput. Sci.} \textbf{2023},
  \emph{3}, 12--24\relax
\mciteBstWouldAddEndPuncttrue
\mciteSetBstMidEndSepPunct{\mcitedefaultmidpunct}
{\mcitedefaultendpunct}{\mcitedefaultseppunct}\relax
\EndOfBibitem
\bibitem[Margraf \latin{et~al.}(2023)Margraf, Jung, Scheurer, and
  Reuter]{Margraf2023}
Margraf,~J.~T.; Jung,~H.; Scheurer,~C.; Reuter,~K. {Exploring Catalytic
  Networks with Machine Learning}. \emph{Nat. Catal.} \textbf{2023}, \emph{6},
  112--121\relax
\mciteBstWouldAddEndPuncttrue
\mciteSetBstMidEndSepPunct{\mcitedefaultmidpunct}
{\mcitedefaultendpunct}{\mcitedefaultseppunct}\relax
\EndOfBibitem
\bibitem[Bergeler \latin{et~al.}(2015)Bergeler, Simm, Proppe, and
  Reiher]{Bergeler2015}
Bergeler,~M.; Simm,~G.~N.; Proppe,~J.; Reiher,~M. {Heuristics-Guided
  Exploration of Reaction Mechanisms}. \emph{J. Chem. Theory Comput.}
  \textbf{2015}, \emph{11}, 5712--5722\relax
\mciteBstWouldAddEndPuncttrue
\mciteSetBstMidEndSepPunct{\mcitedefaultmidpunct}
{\mcitedefaultendpunct}{\mcitedefaultseppunct}\relax
\EndOfBibitem
\bibitem[Simm and Reiher(2017)Simm, and Reiher]{Simm2017}
Simm,~G.~N.; Reiher,~M. {Context-Driven Exploration of Complex Chemical
  Reaction Networks}. \emph{J. Chem. Theory Comput.} \textbf{2017}, \emph{13},
  6108--6119\relax
\mciteBstWouldAddEndPuncttrue
\mciteSetBstMidEndSepPunct{\mcitedefaultmidpunct}
{\mcitedefaultendpunct}{\mcitedefaultseppunct}\relax
\EndOfBibitem
\bibitem[Grimmel and Reiher(2021)Grimmel, and Reiher]{Grimmel2021}
Grimmel,~S.~A.; Reiher,~M. {On the Predictive Power of Chemical Concepts}.
  \emph{Chimia} \textbf{2021}, \emph{75}, 311--311\relax
\mciteBstWouldAddEndPuncttrue
\mciteSetBstMidEndSepPunct{\mcitedefaultmidpunct}
{\mcitedefaultendpunct}{\mcitedefaultseppunct}\relax
\EndOfBibitem
\bibitem[Unsleber \latin{et~al.}(2022)Unsleber, Grimmel, and
  Reiher]{Unsleber2022a}
Unsleber,~J.~P.; Grimmel,~S.~A.; Reiher,~M. {Chemoton 2.0: Autonomous
  Exploration of Chemical Reaction Networks}. \emph{J. Chem. Theory Comput.}
  \textbf{2022}, \emph{18}, 5393--5409\relax
\mciteBstWouldAddEndPuncttrue
\mciteSetBstMidEndSepPunct{\mcitedefaultmidpunct}
{\mcitedefaultendpunct}{\mcitedefaultseppunct}\relax
\EndOfBibitem
\bibitem[Csizi \latin{et~al.}(2024)Csizi, Steiner, and Reiher]{Csizi2024}
Csizi,~K.-S.; Steiner,~M.; Reiher,~M. {Nanoscale Chemical Reaction Exploration
  with a Quantum Magnifying Glass}. \emph{Nat. Commun.} \textbf{2024},
  \emph{15}, 5320\relax
\mciteBstWouldAddEndPuncttrue
\mciteSetBstMidEndSepPunct{\mcitedefaultmidpunct}
{\mcitedefaultendpunct}{\mcitedefaultseppunct}\relax
\EndOfBibitem
\bibitem[Bensberg \latin{et~al.}(2022)Bensberg, T{\"u}rtscher, Unsleber,
  Reiher, and Neugebauer]{Bensberg2022}
Bensberg,~M.; T{\"u}rtscher,~P.~L.; Unsleber,~J.~P.; Reiher,~M.; Neugebauer,~J.
  {Solvation Free Energies in Subsystem Density Functional Theory}. \emph{J.
  Chem. Theory Comput.} \textbf{2022}, \emph{18}, 723--740\relax
\mciteBstWouldAddEndPuncttrue
\mciteSetBstMidEndSepPunct{\mcitedefaultmidpunct}
{\mcitedefaultendpunct}{\mcitedefaultseppunct}\relax
\EndOfBibitem
\bibitem[Brunken and Reiher(2020)Brunken, and Reiher]{Brunken2020}
Brunken,~C.; Reiher,~M. {Self-Parametrizing System-Focused Atomistic Models}.
  \emph{J. Chem. Theory Comput.} \textbf{2020}, \relax
\mciteBstWouldAddEndPunctfalse
\mciteSetBstMidEndSepPunct{\mcitedefaultmidpunct}
{}{\mcitedefaultseppunct}\relax
\EndOfBibitem
\bibitem[Brunken and Reiher(2021)Brunken, and Reiher]{Brunken2021a}
Brunken,~C.; Reiher,~M. {Automated Construction of Quantum--Classical Hybrid
  Models}. \emph{J. Chem. Theory Comput.} \textbf{2021}, \emph{17},
  3797--3813\relax
\mciteBstWouldAddEndPuncttrue
\mciteSetBstMidEndSepPunct{\mcitedefaultmidpunct}
{\mcitedefaultendpunct}{\mcitedefaultseppunct}\relax
\EndOfBibitem
\bibitem[Rumble(2019)]{Rumble2019}
Rumble,~J.~R. \emph{{CRC Handbook of Chemistry and Physics}}; 2019; Vol. 100th
  Edition (Internet Version 2019)\relax
\mciteBstWouldAddEndPuncttrue
\mciteSetBstMidEndSepPunct{\mcitedefaultmidpunct}
{\mcitedefaultendpunct}{\mcitedefaultseppunct}\relax
\EndOfBibitem
\bibitem[Batatia \latin{et~al.}(2022)Batatia, Kovacs, Simm, Ortner, and
  Csanyi]{Batatia2022}
Batatia,~I.; Kovacs,~D.~P.; Simm,~G.; Ortner,~C.; Csanyi,~G. {MACE: Higher
  Order Equivariant Message Passing Neural Networks for Fast and Accurate Force
  Fields}. {Advances in Neural Information Processing Systems}. 2022; pp
  11423--11436\relax
\mciteBstWouldAddEndPuncttrue
\mciteSetBstMidEndSepPunct{\mcitedefaultmidpunct}
{\mcitedefaultendpunct}{\mcitedefaultseppunct}\relax
\EndOfBibitem
\bibitem[Chen and Ong(2022)Chen, and Ong]{Chen2022}
Chen,~C.; Ong,~S.~P. {A Universal Graph Deep Learning Interatomic Potential for
  the Periodic Table}. \emph{Nat. Comput. Sci.} \textbf{2022}, \emph{2},
  718--728\relax
\mciteBstWouldAddEndPuncttrue
\mciteSetBstMidEndSepPunct{\mcitedefaultmidpunct}
{\mcitedefaultendpunct}{\mcitedefaultseppunct}\relax
\EndOfBibitem
\bibitem[Eckhoff and Reiher(2023)Eckhoff, and Reiher]{Eckhoff2023}
Eckhoff,~M.; Reiher,~M. {Lifelong Machine Learning Potentials}. \emph{J. Chem.
  Theory Comput.} \textbf{2023}, \emph{19}, 3509--3525\relax
\mciteBstWouldAddEndPuncttrue
\mciteSetBstMidEndSepPunct{\mcitedefaultmidpunct}
{\mcitedefaultendpunct}{\mcitedefaultseppunct}\relax
\EndOfBibitem
\bibitem[Simm \latin{et~al.}(2020)Simm, T{\"u}rtscher, and Reiher]{Simm2020}
Simm,~G.~N.; T{\"u}rtscher,~P.~L.; Reiher,~M. {Systematic Microsolvation
  Approach with a Cluster-Continuum Scheme and Conformational Sampling}.
  \emph{J. Comput. Chem.} \textbf{2020}, \emph{41}, 1144--1155\relax
\mciteBstWouldAddEndPuncttrue
\mciteSetBstMidEndSepPunct{\mcitedefaultmidpunct}
{\mcitedefaultendpunct}{\mcitedefaultseppunct}\relax
\EndOfBibitem
\bibitem[Steiner \latin{et~al.}(2021)Steiner, Holzknecht, Schauperl, and
  Podewitz]{Steiner2021}
Steiner,~M.; Holzknecht,~T.; Schauperl,~M.; Podewitz,~M. {Quantum Chemical
  Microsolvation by Automated Water Placement}. \emph{Molecules} \textbf{2021},
  \emph{26}, 1793\relax
\mciteBstWouldAddEndPuncttrue
\mciteSetBstMidEndSepPunct{\mcitedefaultmidpunct}
{\mcitedefaultendpunct}{\mcitedefaultseppunct}\relax
\EndOfBibitem
\bibitem[Talmazan and Podewitz(2023)Talmazan, and Podewitz]{Talmazan2023}
Talmazan,~R.~A.; Podewitz,~M. {PyConSolv: A Python Package for Conformer
  Generation of (Metal-Containing) Systems in Explicit Solvent}. \emph{J. Chem.
  Inf. Model.} \textbf{2023}, \emph{63}, 5400--5407\relax
\mciteBstWouldAddEndPuncttrue
\mciteSetBstMidEndSepPunct{\mcitedefaultmidpunct}
{\mcitedefaultendpunct}{\mcitedefaultseppunct}\relax
\EndOfBibitem
\bibitem[Bofill(1994)]{Bofill1994}
Bofill,~J.~M. {Updated Hessian matrix and the restricted step method for
  locating transition structures}. \emph{J. Comput. Chem.} \textbf{1994},
  \emph{15}, 1--11\relax
\mciteBstWouldAddEndPuncttrue
\mciteSetBstMidEndSepPunct{\mcitedefaultmidpunct}
{\mcitedefaultendpunct}{\mcitedefaultseppunct}\relax
\EndOfBibitem
\bibitem[Eyring(1935)]{Eyring1935}
Eyring,~H. {The Activated Complex in Chemical Reactions}. \emph{J. Chem. Phys.}
  \textbf{1935}, \emph{3}, 107--115\relax
\mciteBstWouldAddEndPuncttrue
\mciteSetBstMidEndSepPunct{\mcitedefaultmidpunct}
{\mcitedefaultendpunct}{\mcitedefaultseppunct}\relax
\EndOfBibitem
\bibitem[Swendsen(2019)]{Swendsen2019}
Swendsen,~R.~H. \emph{{An Introduction to Statistical Mechanics and
  Thermodynamics}}; {Oxford Graduate Texts}; 2019\relax
\mciteBstWouldAddEndPuncttrue
\mciteSetBstMidEndSepPunct{\mcitedefaultmidpunct}
{\mcitedefaultendpunct}{\mcitedefaultseppunct}\relax
\EndOfBibitem
\bibitem[Neugebauer \latin{et~al.}(2002)Neugebauer, Reiher, Kind, and
  Hess]{Neugebauer2002}
Neugebauer,~J.; Reiher,~M.; Kind,~C.; Hess,~B.~A. {Quantum Chemical Calculation
  of Vibrational Spectra of Large Molecules---Raman and IR Spectra for
  Buckminsterfullerene}. \emph{J. Comput. Chem.} \textbf{2002}, \emph{23},
  895--910\relax
\mciteBstWouldAddEndPuncttrue
\mciteSetBstMidEndSepPunct{\mcitedefaultmidpunct}
{\mcitedefaultendpunct}{\mcitedefaultseppunct}\relax
\EndOfBibitem
\bibitem[Piccini and Sauer(2014)Piccini, and Sauer]{Piccini2014}
Piccini,~G.; Sauer,~J. {Effect of Anharmonicity on Adsorption Thermodynamics}.
  \emph{J. Chem. Theory Comput.} \textbf{2014}, \emph{10}, 2479--2487\relax
\mciteBstWouldAddEndPuncttrue
\mciteSetBstMidEndSepPunct{\mcitedefaultmidpunct}
{\mcitedefaultendpunct}{\mcitedefaultseppunct}\relax
\EndOfBibitem
\bibitem[Rybicki and Sauer(2022)Rybicki, and Sauer]{Rybicki2022}
Rybicki,~M.; Sauer,~J. {Rigid Body Approximation for the Anharmonic Description
  of Molecule--Surface Vibrations}. \emph{J. Chem. Theory Comput.}
  \textbf{2022}, \emph{18}, 5618--5635\relax
\mciteBstWouldAddEndPuncttrue
\mciteSetBstMidEndSepPunct{\mcitedefaultmidpunct}
{\mcitedefaultendpunct}{\mcitedefaultseppunct}\relax
\EndOfBibitem
\bibitem[Grimme(2012)]{Grimme2012}
Grimme,~S. {Supramolecular Binding Thermodynamics by Dispersion-Corrected
  Density Functional Theory}. \emph{Chem. Eur. J.} \textbf{2012}, \emph{18},
  9955--9964\relax
\mciteBstWouldAddEndPuncttrue
\mciteSetBstMidEndSepPunct{\mcitedefaultmidpunct}
{\mcitedefaultendpunct}{\mcitedefaultseppunct}\relax
\EndOfBibitem
\bibitem[Conquest \latin{et~al.}(2021)Conquest, Roman, Marianov, Kochubei,
  Jiang, and Stampfl]{Conquest2021}
Conquest,~O.~J.; Roman,~T.; Marianov,~A.; Kochubei,~A.; Jiang,~Y.; Stampfl,~C.
  {Calculating Entropies of Large Molecules in Aqueous Phase}. \emph{J. Chem.
  Theory Comput.} \textbf{2021}, \emph{17}, 7753--7771\relax
\mciteBstWouldAddEndPuncttrue
\mciteSetBstMidEndSepPunct{\mcitedefaultmidpunct}
{\mcitedefaultendpunct}{\mcitedefaultseppunct}\relax
\EndOfBibitem
\bibitem[Giesen \latin{et~al.}(1994)Giesen, Cramer, and Truhlar]{Giesen1994}
Giesen,~D.~J.; Cramer,~C.~J.; Truhlar,~D.~G. {Entropic Contributions to Free
  Energies of Solvation}. \emph{J. Phys. Chem.} \textbf{1994}, \emph{98},
  4141--4147\relax
\mciteBstWouldAddEndPuncttrue
\mciteSetBstMidEndSepPunct{\mcitedefaultmidpunct}
{\mcitedefaultendpunct}{\mcitedefaultseppunct}\relax
\EndOfBibitem
\bibitem[Ratkova \latin{et~al.}(2015)Ratkova, Palmer, and Fedorov]{Ratkova2015}
Ratkova,~E.~L.; Palmer,~D.~S.; Fedorov,~M.~V. {Solvation Thermodynamics of
  Organic Molecules by the Molecular Integral Equation Theory: Approaching
  Chemical Accuracy}. \emph{Chem. Rev.} \textbf{2015}, \emph{115},
  6312--6356\relax
\mciteBstWouldAddEndPuncttrue
\mciteSetBstMidEndSepPunct{\mcitedefaultmidpunct}
{\mcitedefaultendpunct}{\mcitedefaultseppunct}\relax
\EndOfBibitem
\bibitem[Besora \latin{et~al.}(2018)Besora, Vidossich, Lled{\'o}s, Ujaque, and
  Maseras]{Besora2018}
Besora,~M.; Vidossich,~P.; Lled{\'o}s,~A.; Ujaque,~G.; Maseras,~F. {Calculation
  of Reaction Free Energies in Solution: A Comparison of Current Approaches}.
  \emph{J. Phys. Chem. A} \textbf{2018}, \emph{122}, 1392--1399\relax
\mciteBstWouldAddEndPuncttrue
\mciteSetBstMidEndSepPunct{\mcitedefaultmidpunct}
{\mcitedefaultendpunct}{\mcitedefaultseppunct}\relax
\EndOfBibitem
\bibitem[Garza(2019)]{Garza2019}
Garza,~A.~J. {Solvation Entropy Made Simple}. \emph{J. Chem. Theory Comput.}
  \textbf{2019}, \emph{15}, 3204--3214\relax
\mciteBstWouldAddEndPuncttrue
\mciteSetBstMidEndSepPunct{\mcitedefaultmidpunct}
{\mcitedefaultendpunct}{\mcitedefaultseppunct}\relax
\EndOfBibitem
\bibitem[Gorges \latin{et~al.}(2022)Gorges, Grimme, Hansen, and
  Pracht]{Gorges2022}
Gorges,~J.; Grimme,~S.; Hansen,~A.; Pracht,~P. {Towards Understanding Solvation
  Effects on the Conformational Entropy of Non-Rigid Molecules}. \emph{Phys.
  Chem. Chem. Phys.} \textbf{2022}, \emph{24}, 12249--12259\relax
\mciteBstWouldAddEndPuncttrue
\mciteSetBstMidEndSepPunct{\mcitedefaultmidpunct}
{\mcitedefaultendpunct}{\mcitedefaultseppunct}\relax
\EndOfBibitem
\bibitem[Pierotti(1976)]{Pierotti1976}
Pierotti,~R.~A. {A Scaled Particle Theory of Aqueous and Nonaqueous Solutions}.
  \emph{Chem. Rev.} \textbf{1976}, \emph{76}, 717--726\relax
\mciteBstWouldAddEndPuncttrue
\mciteSetBstMidEndSepPunct{\mcitedefaultmidpunct}
{\mcitedefaultendpunct}{\mcitedefaultseppunct}\relax
\EndOfBibitem
\bibitem[Weymuth \latin{et~al.}(2024)Weymuth, Unsleber, T{\"u}rtscher, Steiner,
  Sobez, M{\"u}ller, M{\"o}rchen, Klasovita, Grimmel, Eckhoff, Csizi, Bosia,
  Bensberg, and Reiher]{Weymuth2024}
Weymuth,~T.; Unsleber,~J.~P.; T{\"u}rtscher,~P.~L.; Steiner,~M.; Sobez,~J.-G.;
  M{\"u}ller,~C.~H.; M{\"o}rchen,~M.; Klasovita,~V.; Grimmel,~S.~A.;
  Eckhoff,~M.; Csizi,~K.-S.; Bosia,~F.; Bensberg,~M.; Reiher,~M.
  {SCINE---Software for Chemical Interaction Networks}. \emph{J. Chem. Phys.}
  \textbf{2024}, \emph{160}, 222501\relax
\mciteBstWouldAddEndPuncttrue
\mciteSetBstMidEndSepPunct{\mcitedefaultmidpunct}
{\mcitedefaultendpunct}{\mcitedefaultseppunct}\relax
\EndOfBibitem
\bibitem[Bensberg \latin{et~al.}(2023)Bensberg, Brunken, Csizi, Grimmel,
  Gugler, Sobez, Steiner, T{\"u}rtscher, Unsleber, Weymuth, and
  Reiher]{QcscinePuffinAll2024}
Bensberg,~M.; Brunken,~C.; Csizi,~K.-S.; Grimmel,~S.; Gugler,~S.; Sobez,~J.-G.;
  Steiner,~M.; T{\"u}rtscher,~P.~L.; Unsleber,~J.~P.; Weymuth,~T.; Reiher,~M.
  {SCINE Puffin}. 2023; DOI: 10.5281/zenodo.6695461\relax
\mciteBstWouldAddEndPuncttrue
\mciteSetBstMidEndSepPunct{\mcitedefaultmidpunct}
{\mcitedefaultendpunct}{\mcitedefaultseppunct}\relax
\EndOfBibitem
\bibitem[Bensberg \latin{et~al.}(2023)Bensberg, Grimmel, Lang, Simm, Sobez,
  Steiner, T{\"u}rtscher, Unsleber, Weymuth, and
  Reiher]{QcscineChemotonAll2024}
Bensberg,~M.; Grimmel,~S.; Lang,~L.; Simm,~G.~N.; Sobez,~J.-G.; Steiner,~M.;
  T{\"u}rtscher,~P.~L.; Unsleber,~J.~P.; Weymuth,~T.; Reiher,~M. {SCINE
  Chemoton}. 2023; DOI: 10.5281/zenodo.6695583\relax
\mciteBstWouldAddEndPuncttrue
\mciteSetBstMidEndSepPunct{\mcitedefaultmidpunct}
{\mcitedefaultendpunct}{\mcitedefaultseppunct}\relax
\EndOfBibitem
\bibitem[Sobez and Reiher(2020)Sobez, and Reiher]{Sobez2020}
Sobez,~J.-G.; Reiher,~M. {Molassembler: Molecular Graph Construction,
  Modification, and Conformer Generation for Inorganic and Organic Molecules}.
  \emph{J. Chem. Inf. Model.} \textbf{2020}, \emph{60}, 3884--3900\relax
\mciteBstWouldAddEndPuncttrue
\mciteSetBstMidEndSepPunct{\mcitedefaultmidpunct}
{\mcitedefaultendpunct}{\mcitedefaultseppunct}\relax
\EndOfBibitem
\bibitem[Bensberg \latin{et~al.}(2023)Bensberg, Grimmel, Sobez, Steiner,
  Unsleber, and Reiher]{QcscineMolassemblerAll2024}
Bensberg,~M.; Grimmel,~S.; Sobez,~J.-G.; Steiner,~M.; Unsleber,~J.~P.;
  Reiher,~M. {SCINE Molassembler}. 2023; DOI: 10.5281/zenodo.4293554\relax
\mciteBstWouldAddEndPuncttrue
\mciteSetBstMidEndSepPunct{\mcitedefaultmidpunct}
{\mcitedefaultendpunct}{\mcitedefaultseppunct}\relax
\EndOfBibitem
\bibitem[Bensberg \latin{et~al.}(2023)Bensberg, Grimmel, Sobez, Steiner,
  T{\"u}rtscher, Unsleber, and Reiher]{QcscineDatabaseAll2024}
Bensberg,~M.; Grimmel,~S.; Sobez,~J.-G.; Steiner,~M.; T{\"u}rtscher,~P.~L.;
  Unsleber,~J.~P.; Reiher,~M. {SCINE Database}. 2023; DOI:
  10.5281/zenodo.10159610\relax
\mciteBstWouldAddEndPuncttrue
\mciteSetBstMidEndSepPunct{\mcitedefaultmidpunct}
{\mcitedefaultendpunct}{\mcitedefaultseppunct}\relax
\EndOfBibitem
\bibitem[Vaucher and Reiher(2018)Vaucher, and Reiher]{Vaucher2018}
Vaucher,~A.~C.; Reiher,~M. {Minimum Energy Paths and Transition States by Curve
  Optimization}. \emph{J. Chem. Theory Comput.} \textbf{2018}, \emph{14},
  3091--3099\relax
\mciteBstWouldAddEndPuncttrue
\mciteSetBstMidEndSepPunct{\mcitedefaultmidpunct}
{\mcitedefaultendpunct}{\mcitedefaultseppunct}\relax
\EndOfBibitem
\bibitem[Bensberg \latin{et~al.}(2023)Bensberg, Brunken, Csizi, Grimmel,
  Gugler, Sobez, Steiner, T{\"u}rtscher, Unsleber, Vaucher, Weymuth, and
  Reiher]{QcscineReaductAll2024}
Bensberg,~M.; Brunken,~C.; Csizi,~K.-S.; Grimmel,~S.; Gugler,~S.; Sobez,~J.-G.;
  Steiner,~M.; T{\"u}rtscher,~P.~L.; Unsleber,~J.~P.; Vaucher,~A.~C.;
  Weymuth,~T.; Reiher,~M. {SCINE Readuct}. 2023; DOI:
  10.5281/zenodo.3244107\relax
\mciteBstWouldAddEndPuncttrue
\mciteSetBstMidEndSepPunct{\mcitedefaultmidpunct}
{\mcitedefaultendpunct}{\mcitedefaultseppunct}\relax
\EndOfBibitem
\bibitem[Bosia \latin{et~al.}(2023)Bosia, Brunken, Csizi, Sobez, Steiner,
  Unsleber, Weymuth, and Reiher]{QcscineCoreAll2024}
Bosia,~F.; Brunken,~C.; Csizi,~K.-S.; Sobez,~J.-G.; Steiner,~M.;
  Unsleber,~J.~P.; Weymuth,~T.; Reiher,~M. {SCINE Core}. 2023; DOI:
  10.5281/zenodo.3828682\relax
\mciteBstWouldAddEndPuncttrue
\mciteSetBstMidEndSepPunct{\mcitedefaultmidpunct}
{\mcitedefaultendpunct}{\mcitedefaultseppunct}\relax
\EndOfBibitem
\bibitem[Bensberg \latin{et~al.}(2023)Bensberg, Csizi, Grimmel, Sobez, Steiner,
  T{\"u}rtscher, Unsleber, and Reiher]{QcscineXtbAll2024}
Bensberg,~M.; Csizi,~K.-S.; Grimmel,~S.; Sobez,~J.-G.; Steiner,~M.;
  T{\"u}rtscher,~P.~L.; Unsleber,~J.~P.; Reiher,~M. {SCINE Xtb\_wrapper}. 2023;
  DOI: 10.5281/zenodo.5782860\relax
\mciteBstWouldAddEndPuncttrue
\mciteSetBstMidEndSepPunct{\mcitedefaultmidpunct}
{\mcitedefaultendpunct}{\mcitedefaultseppunct}\relax
\EndOfBibitem
\bibitem[Bannwarth \latin{et~al.}(2019)Bannwarth, Ehlert, and
  Grimme]{Bannwarth2019}
Bannwarth,~C.; Ehlert,~S.; Grimme,~S. {GFN2-xTB---An Accurate and Broadly
  Parametrized Self-Consistent Tight-Binding Quantum Chemical Method with
  Multipole Electrostatics and Density-Dependent Dispersion Contributions}.
  \emph{J. Chem. Theory Comput.} \textbf{2019}, \emph{15}, 1652--1671\relax
\mciteBstWouldAddEndPuncttrue
\mciteSetBstMidEndSepPunct{\mcitedefaultmidpunct}
{\mcitedefaultendpunct}{\mcitedefaultseppunct}\relax
\EndOfBibitem
\bibitem[Brunken \latin{et~al.}(2021)Brunken, Csizi, and
  Reiher]{QcscineSwooseAll2024}
Brunken,~C.; Csizi,~K.-S.; Reiher,~M. {SCINE Swoose}. 2021; DOI:
  10.5281/zenodo.5782876\relax
\mciteBstWouldAddEndPuncttrue
\mciteSetBstMidEndSepPunct{\mcitedefaultmidpunct}
{\mcitedefaultendpunct}{\mcitedefaultseppunct}\relax
\EndOfBibitem
\bibitem[Barber \latin{et~al.}(1996)Barber, Dobkin, and Huhdanpaa]{Barber1996}
Barber,~C.~B.; Dobkin,~D.~P.; Huhdanpaa,~H. {The Quickhull Algorithm for Convex
  Hulls}. \emph{ACM Trans. Math. Softw.} \textbf{1996}, \emph{22},
  469--483\relax
\mciteBstWouldAddEndPuncttrue
\mciteSetBstMidEndSepPunct{\mcitedefaultmidpunct}
{\mcitedefaultendpunct}{\mcitedefaultseppunct}\relax
\EndOfBibitem
\bibitem[Balasubramani \latin{et~al.}(2020)Balasubramani, Chen, Coriani,
  Diedenhofen, Frank, Franzke, Furche, Grotjahn, Harding, H{\"a}ttig, Hellweg,
  {Helmich-Paris}, Holzer, Huniar, Kaupp, Marefat~Khah, Karbalaei~Khani,
  M{\"u}ller, Mack, Nguyen, Parker, Perlt, Rappoport, Reiter, Roy, R{\"u}ckert,
  Schmitz, Sierka, Tapavicza, Tew, {van W{\"u}llen}, Voora, Weigend,
  Wody{\'n}ski, and Yu]{Balasubramani2020}
Balasubramani,~S.~G.; Chen,~G.~P.; Coriani,~S.; Diedenhofen,~M.; Frank,~M.~S.;
  Franzke,~Y.~J.; Furche,~F.; Grotjahn,~R.; Harding,~M.~E.; H{\"a}ttig,~C.;
  Hellweg,~A.; {Helmich-Paris},~B.; Holzer,~C.; Huniar,~U.; Kaupp,~M.;
  Marefat~Khah,~A.; Karbalaei~Khani,~S.; M{\"u}ller,~T.; Mack,~F.;
  Nguyen,~B.~D.; Parker,~S.~M.; Perlt,~E.; Rappoport,~D.; Reiter,~K.; Roy,~S.;
  R{\"u}ckert,~M.; Schmitz,~G.; Sierka,~M.; Tapavicza,~E.; Tew,~D.~P.; {van
  W{\"u}llen},~C.; Voora,~V.~K.; Weigend,~F.; Wody{\'n}ski,~A.; Yu,~J.~M.
  {TURBOMOLE: Modular Program Suite for \textit{Ab Initio} Quantum-Chemical and
  Condensed-Matter Simulations}. \emph{J. Chem. Phys.} \textbf{2020},
  \emph{152}, 184107\relax
\mciteBstWouldAddEndPuncttrue
\mciteSetBstMidEndSepPunct{\mcitedefaultmidpunct}
{\mcitedefaultendpunct}{\mcitedefaultseppunct}\relax
\EndOfBibitem
\bibitem[Perdew and Wang(1992)Perdew, and Wang]{Perdew1992}
Perdew,~J.~P.; Wang,~Y. {Accurate and Simple Analytic Representation of the
  Electron-Gas Correlation Energy}. \emph{Phys. Rev. B} \textbf{1992},
  \emph{45}, 13244--13249\relax
\mciteBstWouldAddEndPuncttrue
\mciteSetBstMidEndSepPunct{\mcitedefaultmidpunct}
{\mcitedefaultendpunct}{\mcitedefaultseppunct}\relax
\EndOfBibitem
\bibitem[Perdew \latin{et~al.}(1996)Perdew, Burke, and Ernzerhof]{Perdew1996b}
Perdew,~J.~P.; Burke,~K.; Ernzerhof,~M. {Generalized Gradient Approximation
  Made Simple}. \emph{Phys. Rev. Lett.} \textbf{1996}, \emph{77},
  3865--3868\relax
\mciteBstWouldAddEndPuncttrue
\mciteSetBstMidEndSepPunct{\mcitedefaultmidpunct}
{\mcitedefaultendpunct}{\mcitedefaultseppunct}\relax
\EndOfBibitem
\bibitem[Grimme \latin{et~al.}(2010)Grimme, Antony, Ehrlich, and
  Krieg]{Grimme2010}
Grimme,~S.; Antony,~J.; Ehrlich,~S.; Krieg,~H. {A Consistent and Accurate Ab
  Initio Parametrization of Density Functional Dispersion Correction (DFT-D)
  for the 94 Elements H-Pu}. \emph{J. Chem. Phys.} \textbf{2010}, \emph{132},
  154104\relax
\mciteBstWouldAddEndPuncttrue
\mciteSetBstMidEndSepPunct{\mcitedefaultmidpunct}
{\mcitedefaultendpunct}{\mcitedefaultseppunct}\relax
\EndOfBibitem
\bibitem[Grimme \latin{et~al.}(2011)Grimme, Ehrlich, and Goerigk]{Grimme2011}
Grimme,~S.; Ehrlich,~S.; Goerigk,~L. {Effect of the Damping Function in
  Dispersion Corrected Density Functional Theory}. \emph{J. Comput. Chem.}
  \textbf{2011}, \emph{32}, 1456--1465\relax
\mciteBstWouldAddEndPuncttrue
\mciteSetBstMidEndSepPunct{\mcitedefaultmidpunct}
{\mcitedefaultendpunct}{\mcitedefaultseppunct}\relax
\EndOfBibitem
\bibitem[Peterson \latin{et~al.}(2003)Peterson, Figgen, Goll, Stoll, and
  Dolg]{Peterson2003b}
Peterson,~K.~A.; Figgen,~D.; Goll,~E.; Stoll,~H.; Dolg,~M. {Systematically
  convergent basis sets with relativistic pseudopotentials. II. Small-core
  pseudopotentials and correlation consistent basis sets for the post-d group
  16-18 elements}. \emph{J. Chem. Phys.} \textbf{2003}, \emph{119},
  11113--11123\relax
\mciteBstWouldAddEndPuncttrue
\mciteSetBstMidEndSepPunct{\mcitedefaultmidpunct}
{\mcitedefaultendpunct}{\mcitedefaultseppunct}\relax
\EndOfBibitem
\bibitem[Weigend and Ahlrichs(2005)Weigend, and Ahlrichs]{Weigend2005a}
Weigend,~F.; Ahlrichs,~R. {Balanced basis sets of split valence, triple zeta
  valence and quadruple zeta valence quality for H to Rn: Design and assessment
  of accuracy}. \emph{Phys. Chem. Chem. Phys.} \textbf{2005}, \emph{7},
  3297\relax
\mciteBstWouldAddEndPuncttrue
\mciteSetBstMidEndSepPunct{\mcitedefaultmidpunct}
{\mcitedefaultendpunct}{\mcitedefaultseppunct}\relax
\EndOfBibitem
\bibitem[Neese(2012)]{Neese2012}
Neese,~F. {The ORCA Program System}. \emph{WIREs Comput Mol Sci.}
  \textbf{2012}, \emph{2}, 73--78\relax
\mciteBstWouldAddEndPuncttrue
\mciteSetBstMidEndSepPunct{\mcitedefaultmidpunct}
{\mcitedefaultendpunct}{\mcitedefaultseppunct}\relax
\EndOfBibitem
\bibitem[Neese(2018)]{Neese2018}
Neese,~F. {Software Update: The ORCA Program System, Version 4.0}. \emph{WIREs
  Comput Mol Sci.} \textbf{2018}, \emph{8}, e1327\relax
\mciteBstWouldAddEndPuncttrue
\mciteSetBstMidEndSepPunct{\mcitedefaultmidpunct}
{\mcitedefaultendpunct}{\mcitedefaultseppunct}\relax
\EndOfBibitem
\bibitem[Neese \latin{et~al.}(2020)Neese, Wennmohs, Becker, and
  Riplinger]{Neese2020}
Neese,~F.; Wennmohs,~F.; Becker,~U.; Riplinger,~C. {The ORCA Quantum Chemistry
  Program Package}. \emph{J. Chem. Phys.} \textbf{2020}, \emph{152},
  224108\relax
\mciteBstWouldAddEndPuncttrue
\mciteSetBstMidEndSepPunct{\mcitedefaultmidpunct}
{\mcitedefaultendpunct}{\mcitedefaultseppunct}\relax
\EndOfBibitem
\bibitem[Rappoport and Furche(2010)Rappoport, and Furche]{Rappoport2010a}
Rappoport,~D.; Furche,~F. {Property-optimized Gaussian basis sets for molecular
  response calculations}. \emph{J. Chem. Phys.} \textbf{2010}, \emph{133},
  134105\relax
\mciteBstWouldAddEndPuncttrue
\mciteSetBstMidEndSepPunct{\mcitedefaultmidpunct}
{\mcitedefaultendpunct}{\mcitedefaultseppunct}\relax
\EndOfBibitem
\bibitem[{Garcia-Rat{\'e}s} and Neese(2020){Garcia-Rat{\'e}s}, and
  Neese]{Garcia-Rates2020}
{Garcia-Rat{\'e}s},~M.; Neese,~F. {Effect of the Solute Cavity on the Solvation
  Energy and Its Derivatives within the Framework of the Gaussian Charge
  Scheme}. \emph{J. Comput. Chem.} \textbf{2020}, \emph{41}, 922--939\relax
\mciteBstWouldAddEndPuncttrue
\mciteSetBstMidEndSepPunct{\mcitedefaultmidpunct}
{\mcitedefaultendpunct}{\mcitedefaultseppunct}\relax
\EndOfBibitem
\bibitem[T{\"u}rtscher and Reiher(2023)T{\"u}rtscher, and
  Reiher]{Turtscher2023}
T{\"u}rtscher,~P.~L.; Reiher,~M. {Pathfinder-Navigating and Analyzing Chemical
  Reaction Networks with an Efficient Graph-Based Approach}. \emph{J. Chem.
  Inf. Model.} \textbf{2023}, \emph{63}, 147--160\relax
\mciteBstWouldAddEndPuncttrue
\mciteSetBstMidEndSepPunct{\mcitedefaultmidpunct}
{\mcitedefaultendpunct}{\mcitedefaultseppunct}\relax
\EndOfBibitem
\bibitem[Zavitsas \latin{et~al.}(1970)Zavitsas, Coffiner, Wiseman, and
  Zavitsas]{Zavitsas1970}
Zavitsas,~A.~A.; Coffiner,~M.; Wiseman,~T.; Zavitsas,~L.~R. {Reversible
  Hydration of Formaldehyde. Thermodynamic Parameters}. \emph{J. Phys. Chem.}
  \textbf{1970}, \emph{74}, 2746--2750\relax
\mciteBstWouldAddEndPuncttrue
\mciteSetBstMidEndSepPunct{\mcitedefaultmidpunct}
{\mcitedefaultendpunct}{\mcitedefaultseppunct}\relax
\EndOfBibitem
\bibitem[Williams \latin{et~al.}(1983)Williams, Spangler, Femec, Maggiora, and
  Schowen]{Williams1983}
Williams,~I.~H.; Spangler,~D.; Femec,~D.~A.; Maggiora,~G.~M.; Schowen,~R.~L.
  {Theoretical Models for Solvation and Catalysis in Carbonyl Addition}.
  \emph{J. Am. Chem. Soc.} \textbf{1983}, \emph{105}, 31--40\relax
\mciteBstWouldAddEndPuncttrue
\mciteSetBstMidEndSepPunct{\mcitedefaultmidpunct}
{\mcitedefaultendpunct}{\mcitedefaultseppunct}\relax
\EndOfBibitem
\bibitem[Wolfe \latin{et~al.}(1995)Wolfe, Kim, Yang, Weinberg, and
  Shi]{Wolfe1995}
Wolfe,~S.; Kim,~C.-K.; Yang,~K.; Weinberg,~N.; Shi,~Z. {Hydration of the
  Carbonyl Group. A Theoretical Study of the Cooperative Mechanism}. \emph{J.
  Am. Chem. Soc.} \textbf{1995}, \emph{117}, 4240--4260\relax
\mciteBstWouldAddEndPuncttrue
\mciteSetBstMidEndSepPunct{\mcitedefaultmidpunct}
{\mcitedefaultendpunct}{\mcitedefaultseppunct}\relax
\EndOfBibitem
\bibitem[Zhang and Kim(2008)Zhang, and Kim]{Zhang2008}
Zhang,~H.; Kim,~C.-K. {Hydration of Formaldehyde in Water: Insight from ONIOM
  Study}. \emph{Bull. Korean Chem. Soc.} \textbf{2008}, \emph{29},
  2528--2530\relax
\mciteBstWouldAddEndPuncttrue
\mciteSetBstMidEndSepPunct{\mcitedefaultmidpunct}
{\mcitedefaultendpunct}{\mcitedefaultseppunct}\relax
\EndOfBibitem
\bibitem[Inaba(2015)]{Inaba2015}
Inaba,~S. {Theoretical Study of Decomposition of Methanediol in Aqueous
  Solution}. \emph{J. Phys. Chem. A} \textbf{2015}, \emph{119},
  5816--5825\relax
\mciteBstWouldAddEndPuncttrue
\mciteSetBstMidEndSepPunct{\mcitedefaultmidpunct}
{\mcitedefaultendpunct}{\mcitedefaultseppunct}\relax
\EndOfBibitem
\bibitem[Wang \latin{et~al.}(2023)Wang, Chen, Liu, Huang, and Jiang]{Wang2023}
Wang,~C.; Chen,~X.; Liu,~Y.; Huang,~T.; Jiang,~S. {Theoretical Study of the
  Gas-Phase Hydrolysis of Formaldehyde to Produce Methanediol and Its
  Implication to New Particle Formation}. \emph{ACS Omega} \textbf{2023},
  \emph{8}, 15467--15478\relax
\mciteBstWouldAddEndPuncttrue
\mciteSetBstMidEndSepPunct{\mcitedefaultmidpunct}
{\mcitedefaultendpunct}{\mcitedefaultseppunct}\relax
\EndOfBibitem
\bibitem[Winkelman \latin{et~al.}(2002)Winkelman, Voorwinde, Ottens,
  Beenackers, and Janssen]{Winkelman2002}
Winkelman,~J. G.~M.; Voorwinde,~O.~K.; Ottens,~M.; Beenackers,~A. A. C.~M.;
  Janssen,~L. P. B.~M. {Kinetics and Chemical Equilibrium of the Hydration of
  Formaldehyde}. \emph{Chem. Eng. Sci.} \textbf{2002}, \emph{57},
  4067--4076\relax
\mciteBstWouldAddEndPuncttrue
\mciteSetBstMidEndSepPunct{\mcitedefaultmidpunct}
{\mcitedefaultendpunct}{\mcitedefaultseppunct}\relax
\EndOfBibitem
\bibitem[Harvey \latin{et~al.}(2019)Harvey, Himo, Maseras, and
  Perrin]{Harvey2019}
Harvey,~J.~N.; Himo,~F.; Maseras,~F.; Perrin,~L. {Scope and Challenge of
  Computational Methods for Studying Mechanism and Reactivity in Homogeneous
  Catalysis}. \emph{ACS Catal.} \textbf{2019}, \emph{9}, 6803--6813\relax
\mciteBstWouldAddEndPuncttrue
\mciteSetBstMidEndSepPunct{\mcitedefaultmidpunct}
{\mcitedefaultendpunct}{\mcitedefaultseppunct}\relax
\EndOfBibitem
\bibitem[Tantillo(2022)]{Tantillo2022}
Tantillo,~D.~J. {Portable Models for Entropy Effects on Kinetic Selectivity}.
  \emph{J. Am. Chem. Soc.} \textbf{2022}, \emph{144}, 13996--14004\relax
\mciteBstWouldAddEndPuncttrue
\mciteSetBstMidEndSepPunct{\mcitedefaultmidpunct}
{\mcitedefaultendpunct}{\mcitedefaultseppunct}\relax
\EndOfBibitem
\bibitem[Cheong \latin{et~al.}(2024)Cheong, Tipp, Eikerling, and
  Kowalski]{Cheong2024}
Cheong,~O.; Tipp,~F.~P.; Eikerling,~M.~H.; Kowalski,~P.~M. {Entropy Effects on
  Reactive Processes at Metal-Solvent Interfaces}. \emph{J. Phys. Chem. C}
  \textbf{2024}, \emph{128}, 7892--7902\relax
\mciteBstWouldAddEndPuncttrue
\mciteSetBstMidEndSepPunct{\mcitedefaultmidpunct}
{\mcitedefaultendpunct}{\mcitedefaultseppunct}\relax
\EndOfBibitem
\bibitem[Bensberg and Reiher(2023)Bensberg, and Reiher]{Bensberg2023}
Bensberg,~M.; Reiher,~M. {Concentration-Flux-Steered Mechanism Exploration with
  an Organocatalysis Application}. \emph{Isr. J. Chem.} \textbf{2023},
  \emph{n/a}, e202200123\relax
\mciteBstWouldAddEndPuncttrue
\mciteSetBstMidEndSepPunct{\mcitedefaultmidpunct}
{\mcitedefaultendpunct}{\mcitedefaultseppunct}\relax
\EndOfBibitem
\bibitem[Bensberg and Reiher(2024)Bensberg, and Reiher]{Bensberg2024}
Bensberg,~M.; Reiher,~M. {Uncertainty-Aware First-Principles Exploration of
  Chemical Reaction Networks}. \emph{J. Phys. Chem. A} \textbf{2024},
  \emph{128}, 4532--4547\relax
\mciteBstWouldAddEndPuncttrue
\mciteSetBstMidEndSepPunct{\mcitedefaultmidpunct}
{\mcitedefaultendpunct}{\mcitedefaultseppunct}\relax
\EndOfBibitem
\bibitem[Simm and Reiher(2018)Simm, and Reiher]{Simm2018}
Simm,~G.~N.; Reiher,~M. {Error-Controlled Exploration of Chemical Reaction
  Networks with Gaussian Processes}. \emph{J. Chem. Theory Comput.}
  \textbf{2018}, \emph{14}, 5238--5248\relax
\mciteBstWouldAddEndPuncttrue
\mciteSetBstMidEndSepPunct{\mcitedefaultmidpunct}
{\mcitedefaultendpunct}{\mcitedefaultseppunct}\relax
\EndOfBibitem
\bibitem[Bond \latin{et~al.}(2012)Bond, Goslan, Parsons, and
  Jefferson]{Bond2012}
Bond,~T.; Goslan,~E.~H.; Parsons,~S.~A.; Jefferson,~B. {A Critical Review of
  Trihalomethane and Haloacetic Acid Formation from Natural Organic Matter
  Surrogates}. \emph{Environ. Technol. Rev.} \textbf{2012}, \emph{1},
  93--113\relax
\mciteBstWouldAddEndPuncttrue
\mciteSetBstMidEndSepPunct{\mcitedefaultmidpunct}
{\mcitedefaultendpunct}{\mcitedefaultseppunct}\relax
\EndOfBibitem
\bibitem[Prasse \latin{et~al.}(2020)Prasse, {von Gunten}, and
  Sedlak]{Prasse2020}
Prasse,~C.; {von Gunten},~U.; Sedlak,~D.~L. {Chlorination of Phenols Revisited:
  Unexpected Formation of $\alpha$,$\beta$-Unsaturated C4-Dicarbonyl Ring
  Cleavage Products}. \emph{Environ. Sci. Technol.} \textbf{2020}, \emph{54},
  826--834\relax
\mciteBstWouldAddEndPuncttrue
\mciteSetBstMidEndSepPunct{\mcitedefaultmidpunct}
{\mcitedefaultendpunct}{\mcitedefaultseppunct}\relax
\EndOfBibitem
\bibitem[Mazur and Lebedev(2022)Mazur, and Lebedev]{Mazur2022}
Mazur,~D.~M.; Lebedev,~A.~T. {Transformation of Organic Compounds during Water
  Chlorination/Bromination: Formation Pathways for Disinfection By-Products (A
  Review)}. \emph{J. Anal. Chem.} \textbf{2022}, \emph{77}, 1705--1728\relax
\mciteBstWouldAddEndPuncttrue
\mciteSetBstMidEndSepPunct{\mcitedefaultmidpunct}
{\mcitedefaultendpunct}{\mcitedefaultseppunct}\relax
\EndOfBibitem
\bibitem[Gallard and {von Gunten}(2002)Gallard, and {von Gunten}]{Gallard2002}
Gallard,~H.; {von Gunten},~U. {Chlorination of Phenols:\, Kinetics and
  Formation of Chloroform}. \emph{Environ. Sci. Technol.} \textbf{2002},
  \emph{36}, 884--890\relax
\mciteBstWouldAddEndPuncttrue
\mciteSetBstMidEndSepPunct{\mcitedefaultmidpunct}
{\mcitedefaultendpunct}{\mcitedefaultseppunct}\relax
\EndOfBibitem
\bibitem[Leung \latin{et~al.}(2007)Leung, Nielsen, and Kurtz]{Leung2007}
Leung,~K.; Nielsen,~I. M.~B.; Kurtz,~I. {Ab Initio Molecular Dynamics Study of
  Carbon Dioxide and Bicarbonate Hydration and the Nucleophilic Attack of
  Hydroxide on CO2}. \emph{J. Phys. Chem. B} \textbf{2007}, \emph{111},
  4453--4459\relax
\mciteBstWouldAddEndPuncttrue
\mciteSetBstMidEndSepPunct{\mcitedefaultmidpunct}
{\mcitedefaultendpunct}{\mcitedefaultseppunct}\relax
\EndOfBibitem
\bibitem[Kumar \latin{et~al.}(2009)Kumar, Kalinichev, and
  Kirkpatrick]{Kumar2009}
Kumar,~P.~P.; Kalinichev,~A.~G.; Kirkpatrick,~R.~J. {Hydrogen-Bonding Structure
  and Dynamics of Aqueous Carbonate Species from Car-Parrinello Molecular
  Dynamics Simulations}. \emph{J. Phys. Chem. B} \textbf{2009}, \emph{113},
  794--802\relax
\mciteBstWouldAddEndPuncttrue
\mciteSetBstMidEndSepPunct{\mcitedefaultmidpunct}
{\mcitedefaultendpunct}{\mcitedefaultseppunct}\relax
\EndOfBibitem
\bibitem[Stirling and P{\'a}pai(2010)Stirling, and P{\'a}pai]{Stirling2010}
Stirling,~A.; P{\'a}pai,~I. {H2CO3 Forms via HCO3- in Water}. \emph{J. Phys.
  Chem. B} \textbf{2010}, \emph{114}, 16854--16859\relax
\mciteBstWouldAddEndPuncttrue
\mciteSetBstMidEndSepPunct{\mcitedefaultmidpunct}
{\mcitedefaultendpunct}{\mcitedefaultseppunct}\relax
\EndOfBibitem
\bibitem[Gallet \latin{et~al.}(2012)Gallet, Pietrucci, and
  Andreoni]{Gallet2012}
Gallet,~G.~A.; Pietrucci,~F.; Andreoni,~W. {Bridging Static and Dynamical
  Descriptions of Chemical Reactions: An Ab Initio Study of CO2 Interacting
  with Water Molecules}. \emph{J. Chem. Theory Comput.} \textbf{2012},
  \emph{8}, 4029--4039\relax
\mciteBstWouldAddEndPuncttrue
\mciteSetBstMidEndSepPunct{\mcitedefaultmidpunct}
{\mcitedefaultendpunct}{\mcitedefaultseppunct}\relax
\EndOfBibitem
\bibitem[Martirez and Carter(2023)Martirez, and Carter]{Martirez2023}
Martirez,~J. M.~P.; Carter,~E.~A. {Solvent Dynamics Are Critical to
  Understanding Carbon Dioxide Dissolution and Hydration in Water}. \emph{J.
  Am. Chem. Soc.} \textbf{2023}, \emph{145}, 12561--12575\relax
\mciteBstWouldAddEndPuncttrue
\mciteSetBstMidEndSepPunct{\mcitedefaultmidpunct}
{\mcitedefaultendpunct}{\mcitedefaultseppunct}\relax
\EndOfBibitem
\bibitem[Nguyen \latin{et~al.}(1997)Nguyen, Raspoet, Vanquickenborne, and
  Duijnen]{Nguyen1997}
Nguyen,~M.~T.; Raspoet,~G.; Vanquickenborne,~L.~G.; Duijnen,~P. T.~V. {How Many
  Water Molecules Are Actively Involved in the Neutral Hydration of Carbon
  Dioxide?} \emph{J. Phys. Chem. A} \textbf{1997}, \emph{101}, 7379--7388\relax
\mciteBstWouldAddEndPuncttrue
\mciteSetBstMidEndSepPunct{\mcitedefaultmidpunct}
{\mcitedefaultendpunct}{\mcitedefaultseppunct}\relax
\EndOfBibitem
\bibitem[Nguyen \latin{et~al.}(2008)Nguyen, Matus, Jackson, Ngan, Rustad, and
  Dixon]{Nguyen2008}
Nguyen,~M.~T.; Matus,~M.~H.; Jackson,~V.~E.; Ngan,~V.~T.; Rustad,~J.~R.;
  Dixon,~D.~A. {Mechanism of the Hydration of Carbon Dioxide: Direct
  Participation of H2O versus Microsolvation}. \emph{J. Phys. Chem. A}
  \textbf{2008}, \emph{112}, 10386--10398\relax
\mciteBstWouldAddEndPuncttrue
\mciteSetBstMidEndSepPunct{\mcitedefaultmidpunct}
{\mcitedefaultendpunct}{\mcitedefaultseppunct}\relax
\EndOfBibitem
\bibitem[Wang and Cao(2013)Wang, and Cao]{Wang2013}
Wang,~B.; Cao,~Z. {How Water Molecules Modulate the Hydration of CO2 in Water
  Solution: Insight from the Cluster-Continuum Model Calculations}. \emph{J.
  Comput. Chem.} \textbf{2013}, \emph{34}, 372--378\relax
\mciteBstWouldAddEndPuncttrue
\mciteSetBstMidEndSepPunct{\mcitedefaultmidpunct}
{\mcitedefaultendpunct}{\mcitedefaultseppunct}\relax
\EndOfBibitem
\bibitem[Adamczyk \latin{et~al.}(2009)Adamczyk, {Pr{\'e}mont-Schwarz}, Pines,
  Pines, and Nibbering]{Adamczyk2009}
Adamczyk,~K.; {Pr{\'e}mont-Schwarz},~M.; Pines,~D.; Pines,~E.; Nibbering,~E.
  T.~J. {Real-Time Observation of Carbonic Acid Formation in Aqueous Solution}.
  \emph{Science} \textbf{2009}, \emph{326}, 1690--1694\relax
\mciteBstWouldAddEndPuncttrue
\mciteSetBstMidEndSepPunct{\mcitedefaultmidpunct}
{\mcitedefaultendpunct}{\mcitedefaultseppunct}\relax
\EndOfBibitem
\bibitem[Wang \latin{et~al.}(2010)Wang, Conway, Burns, McCann, and
  Maeder]{Wang2010}
Wang,~X.; Conway,~W.; Burns,~R.; McCann,~N.; Maeder,~M. {Comprehensive Study of
  the Hydration and Dehydration Reactions of Carbon Dioxide in Aqueous
  Solution}. \emph{J. Phys. Chem. A} \textbf{2010}, \emph{114},
  1734--1740\relax
\mciteBstWouldAddEndPuncttrue
\mciteSetBstMidEndSepPunct{\mcitedefaultmidpunct}
{\mcitedefaultendpunct}{\mcitedefaultseppunct}\relax
\EndOfBibitem
\bibitem[England \latin{et~al.}(2011)England, Duffin, Schwartz, Uejio,
  Prendergast, and Saykally]{England2011}
England,~A.~H.; Duffin,~A.~M.; Schwartz,~C.~P.; Uejio,~J.~S.; Prendergast,~D.;
  Saykally,~R.~J. {On the Hydration and Hydrolysis of Carbon Dioxide}.
  \emph{Chem. Phys. Lett.} \textbf{2011}, \emph{514}, 187--195\relax
\mciteBstWouldAddEndPuncttrue
\mciteSetBstMidEndSepPunct{\mcitedefaultmidpunct}
{\mcitedefaultendpunct}{\mcitedefaultseppunct}\relax
\EndOfBibitem
\bibitem[H{\"o}nisch \latin{et~al.}(2012)H{\"o}nisch, Ridgwell, Schmidt,
  Thomas, Gibbs, Sluijs, Zeebe, Kump, Martindale, Greene, Kiessling, Ries,
  Zachos, Royer, Barker, Marchitto, Moyer, Pelejero, Ziveri, Foster, and
  Williams]{Honisch2012}
H{\"o}nisch,~B.; Ridgwell,~A.; Schmidt,~D.~N.; Thomas,~E.; Gibbs,~S.~J.;
  Sluijs,~A.; Zeebe,~R.; Kump,~L.; Martindale,~R.~C.; Greene,~S.~E.;
  Kiessling,~W.; Ries,~J.; Zachos,~J.~C.; Royer,~D.~L.; Barker,~S.;
  Marchitto,~T.~M.; Moyer,~R.; Pelejero,~C.; Ziveri,~P.; Foster,~G.~L.;
  Williams,~B. {The Geological Record of Ocean Acidification}. \emph{Science}
  \textbf{2012}, \emph{335}, 1058--1063\relax
\mciteBstWouldAddEndPuncttrue
\mciteSetBstMidEndSepPunct{\mcitedefaultmidpunct}
{\mcitedefaultendpunct}{\mcitedefaultseppunct}\relax
\EndOfBibitem
\bibitem[T{\"u}rtscher and Reiher(2024)T{\"u}rtscher, and
  Reiher]{kingfisher2024}
T{\"u}rtscher,~P.~L.; Reiher,~M. {Data Set for the Journal Article "Automated
  Microsolvation for Minimum Energy Path Construction in Solution"}. 2024;
  DOI:~10.5281/zenodo.13819507\relax
\mciteBstWouldAddEndPuncttrue
\mciteSetBstMidEndSepPunct{\mcitedefaultmidpunct}
{\mcitedefaultendpunct}{\mcitedefaultseppunct}\relax
\EndOfBibitem
\end{mcitethebibliography}
\providecommand{\latin}[1]{#1}
\makeatletter
\providecommand{\doi}
  {\begingroup\let\do\@makeother\dospecials
  \catcode`\{=1 \catcode`\}=2 \doi@aux}
\providecommand{\doi@aux}[1]{\endgroup\texttt{#1}}
\makeatother
\providecommand*\mcitethebibliography{\thebibliography}
\csname @ifundefined\endcsname{endmcitethebibliography}
  {\let\endmcitethebibliography\endthebibliography}{}

%

\end{document}